\documentclass[11pt]{article}
\usepackage{amssymb,amsmath,amsfonts}
\usepackage{graphicx}
\usepackage{graphics}
\usepackage{epsfig}

\makeatletter
\@addtoreset{equation}{section}

\makeatother

\begin{document}

\title{Fermionic Casimir densities for a uniformly accelerating mirror \\
in the Fulling-Rindler vacuum}
\author{A. A. Saharian$^{1}$\thanks{%
E-mail: saharian@ysu.am},\thinspace\ L. Sh. Grigoryan$^{2}$, V. Kh. Kotanjyan%
$^{1,2}$ \\
%EndAName
\\
\textit{$^1$Institute of Physics, Yerevan State University,}\\
\textit{1 Alex Manoogian Street, 0025 Yerevan, Armenia} \vspace{0.3cm}\\
\textit{$^2$Institute of Applied Problems of Physics NAS RA,}\\
\textit{25 Hrachya Nersissyan Street, 0014 Yerevan, Armenia}}
\maketitle

\begin{abstract}
We investigate the local characteristics of the Fulling-Rindler vacuum for a
massive Dirac field induced by a planar boundary moving with constant proper
acceleration in $(D+1)$-dimensional flat spacetime. On the boundary, the
field operator obeys the bag boundary condition. The boundary divides the
right Rindler wedge into two separate regions, called RL and RR regions. In
both these regions, the fermion condensate and the vacuum expectation value
(VEV) of the energy-momentum tensor are decomposed into two contributions.
The first one presents the VEVs in the Fulling-Rindler vacuum when the
boundary is absent and the second one is the boundary-induced contribution.
For points away from the boundary, the renormalization is reduced to the one
for the boundary-free geometry. The total VEVs are dominated by the
boundary-free parts near the Rindler horizon and by the boundary-induced
parts in the region near the boundary. For a massive field the boundary-free
contributions in the fermion condensate and the vacuum energy density and
effective pressures are negative everywhere. The boundary-induced
contributions in the fermion condensate and the energy density are positive
in the RL region and negative in the RR region. For a massless field the
fermion condensate vanishes in spatial dimensions $D\geq 2$, while the VEV
of the energy-momentum tensor is different from zero. This behavior
contrasts with that of the VEVs in the Minkowski vacuum for the geometry of
a boundary at rest relative to an inertial observer. In the latter case, the
fermion condensate for a massless field is nonzero, while the VEV of the
energy-momentum tensor becomes zero. The obtained results are used to
investigate the VEV of the fermionic energy-momentum tensor in weak
gravitational fields and background geometries that are conformally related
to Rindler spacetime.
\end{abstract}

\bigskip


\section{Introduction}

The Rindler coordinates $x^{\mu }=(\tau ,\rho ,x^{2},\ldots ,x^{D})$ in $%
(D+1)$-dimensional flat spacetime, with the variation range $\tau \in
(-\infty ,+\infty )$, $\rho \in \lbrack 0,\infty )$, $x^{i}\in (-\infty
,+\infty )$, $i=2,\ldots ,D$, realize the reference frame of uniformly
accelerated observers. They are related to the Minkowskian coordinates $%
x^{\prime \mu }=(x^{\prime 0}=t,x^{\prime 1},x^{\prime 2},\ldots ,x^{\prime
D})$ of inertial observers by the coordinate transformation%
\begin{equation}
t=\rho \sinh \tau ,\;x^{1}=\rho \cosh \tau ,\;x^{\prime
i}=x^{i},\;i=2,\ldots D.  \label{trans}
\end{equation}%
With this transformation, the line element in the Rindler coordinates reads%
\begin{equation}
ds_{\mathrm{R}}^{2}=g_{\mu \nu }dx^{\mu }dx^{\nu }=\rho ^{2}d\tau ^{2}-d\rho
^{2}-d\mathbf{x}^{2},  \label{ds2}
\end{equation}%
where $\mathbf{x}=(x^{2},x^{3},\ldots ,x^{D})$. The worldlines $\rho =%
\mathrm{const}$ and $\mathbf{x}=\mathrm{const}$ describe a uniformly
accelerating observer following the trajectory parallel to the $x^{1}$ axis
and having a proper acceleration $1/\rho $. The proper time $\tau _{\mathrm{p%
}}$ of the observer is expressed in terms of the dimensionless time
coordinate $\tau $ as $\tau _{\mathrm{p}}=\rho \tau $. The hyperplanes $%
x^{1}=\pm t$ are horizons for Rindler observers. They divide the Minkowski
spacetime into four wedges. For the coordinates defined in (\ref{trans}) one
has $x^{1}>|t|$ and they correspond to the so called Right (R) region. The
coordinate transformation in the wedge $x^{1}<-|t|$ (Left (L) region) is
given by (\ref{trans}) with $x^{1}=-\rho \cosh \tau $. The remaining wedges $%
t>|x^{1}|$ and $t<-|x^{1}|$ present the Future (F) and Past (P) regions of
the Minkowski spacetime. The corresponding transformations read $t=\pm \rho
\cosh \tau $,$\;x^{1}=\rho \sinh \tau $.

The interest to the Rindler spacetime, as a background geometry in quantum
field theory, is motivated by a number of reasons. First of all, the Rindler
geometry is a classic example for discussing fundamental issues such as the
observer dependence of the concepts of vacuum and particles. The Rinlder
metric tensor is static and the canonical quantization of a given field can
be realized by using the modes of a given energy in both Rindler and
Minkowskian coordinates. Expanding the field operator over those modes, the
vacuum and particle states are constructed following the standard procedure.
It turns out that the vacuum states constructed on the basis of Rindlerian
and Minkowskian modes (Fulling-Rinlder and Minkowski vacua, respectively)
are different (see, for example, \cite{Birr82,Cris08,Lang05}). An observer
moving with constant acceleration registers Rindler quanta in the Minkowski
vacuum (the Unruh effect \cite{Full73,Davi75,Unru76}), and an inertial
observer registers particles in the Fulling-Rindler vacuum. The simplest
example of the system registering particles is the Unruh-de Witt point like
detector \cite{Birr82}. An important difference between Rindler observers
and inertial observers is the presence of an event horizon related to the
accelerated motion. The presence of the horizon introduces new qualitative
features in quantum field-theoretical effects. The relative simplicity of
the Rindler geometry allows to obtain closed analytical results for physical
characteristics describing the state of a quantum field. This helps to
understand the consequences of the existence of a horizon in more complex
situations in curved space-time, including those for de Sitter and black
hole geometries. An interesting quantum phenomena is the entanglement
between the states in the regions separated by horizons. In the Rindler
geometry the worldlines of uniformly accelerating observers are located in
the right and left wedges. For those observers, the Minkowski vacuum is an
entangled state that connects the states in separate wedges (see, for
example, \cite{Fuen05}-\cite{Yan22} and references therein).

An intriguing macroscopic manifestation of the non-trivial physical
properties of the quantum vacuum is the Casimir effect (for reviews see \cite%
{Most97}-\cite{Casi11}). It is a boundary-induced phenomenon caused by the
influence of boundaries on quantum fluctuations of fields. The effect has
been thoroughly investigated for both bosonic and fermionic fields and for
various geometries of boundaries and background spacetime. A series of
high-precision experiments confirm the Casimir effect for electromagnetic
field in problems involving conducting and dielectric boundaries. An
interesting topic in the investigations on the Casimir effect is the
dependence of the expectation values of physical quantities on the chosen
vacuum state. In the present paper we study the Casimir contributions in the
fermion condensate and the mean energy-momentum tensor for Dirac field,
induced by a planar boundary moving with constant proper acceleration
through the Fulling-Rindler vacuum state (quantum fermionic fields in
Rindler spacetime have been considered in various aspects in the literature,
see, e.g., Refs. \cite{Cand78}-\cite{Bell23}). Similar studies have been
conducted previously in Refs. \cite{Cand77}-\cite{Saha06} for scalar and
electromagnetic fields in the geometry of a single and two parallel planar
boundaries. Note that the problems of the Fulling-Rindler vacuum
polarization by moving boundaries are closely related to the investigation
of the Casimir effect in weak gravitational fields (see, for example, Refs.
\cite{Bimo06}-\cite{BezeV24}).

The organization of the paper is as follows. The problem setup and the Dirac
modes in the RR and RL regions, separated by a mirror, are presented in the
next section. In Section \ref{sec:FC} we investigate the fermion condensate
in those regions. The boundary-induced contributions are explicitly
separated and their behavior in asymptotic regions of the parameters is
studied. The corresponding investigation for the expectation value of the
energy-momentum tensor is given in Section \ref{sec:EMT}. The main results
are summarized in Section \ref{sec:Conc}. In Appendix \ref{sec:AppNI} we
present the details of the evaluation of the normalization integrals in the
RR and RL regions. In Appendix \ref{sec:AppSum}, by using the generalized
Abel-Plana formula, a summation formula is derived for the series appearing
in the expressions of the expectation values in the RR-region. The proof of
the identities, used for the separation of the boundary-induced
contributions in the expectation values for the RL-region, is presented in
Appendix \ref{sec:AppIdent}.

\section{Problem setup and fermionic modes}

\label{sec:Modes}

\subsection{Problem setup}

Consider a fermionic field $\psi (x^{\mu })$ in background of spacetime
described by the line element (\ref{ds2}). The $(D+1)$-bein fields will be
taken in the form $e_{(0)}^{\mu }=\delta _{0}^{\mu }/\rho $, $e_{(b)}^{\mu
}=\delta _{b}^{\mu }$, $b=1,2,\ldots $, and for the covariant components of
the metric tensor, $g_{\mu \nu }=\eta _{ab}e_{\mu }^{(a)}e_{\nu }^{(b)}$,
one has $g_{\mu \nu }=\mathrm{diag}(\rho ^{2},-1,\ldots ,-1)$. Here, $\eta
_{ab}$ is the Minkowskian metric tensor in Cartesian coordinates with $%
a,b=0,1,\ldots ,D$. The field operator obeys the Dirac equation
\begin{equation}
\left( i\gamma ^{\mu }\nabla _{\mu }-m\right) \psi =0,\;\nabla _{\mu
}=\partial _{\mu }+\Gamma _{\mu },  \label{eom}
\end{equation}%
with the spin connection $\Gamma _{\mu }$. For the $N\times N$ Dirac
matrices in the coordinates $x^{\mu }$ we have the relation $\gamma ^{\mu
}=e_{(b)}^{\mu }\gamma ^{(b)}$, where $\gamma ^{(b)}$ are the flat spacetime
gamma matrices in Cartesian coordinates (for the construction of the
matrices $\gamma ^{(b)}$ in general number of the spacetime dimension see,
e.g., \cite{Shim85,Park09}). They obey the anticommutation relation $\left\{
\gamma ^{\mu },\gamma ^{\nu }\right\} =2g^{\mu \nu }$ and we will assume
that the field $\psi (x^{\mu })$ realizes the irreducible representation of
this Clifford algebra. For odd values of spatial dimension $D$, the
irreducible representation is unique (up to a similarity transformation).
For even $D$ one has two inequivalent irreducible representations. The
dimension of the Dirac matrices in the irreducible representation is given
by $N=2^{[(D+1)/2]}$, where $[\cdots ]$ stands for the integer part of the
enclosed expression. Denoting the covariant derivative for vector fields by $%
;$, the spin connection is expressed as $\Gamma _{\mu }=\gamma ^{(a)}\gamma
^{(b)}e_{(a)}^{\nu }e_{(b)\nu ;\mu }/4$.

\begin{figure}[tbph]
\begin{center}
\epsfig{figure=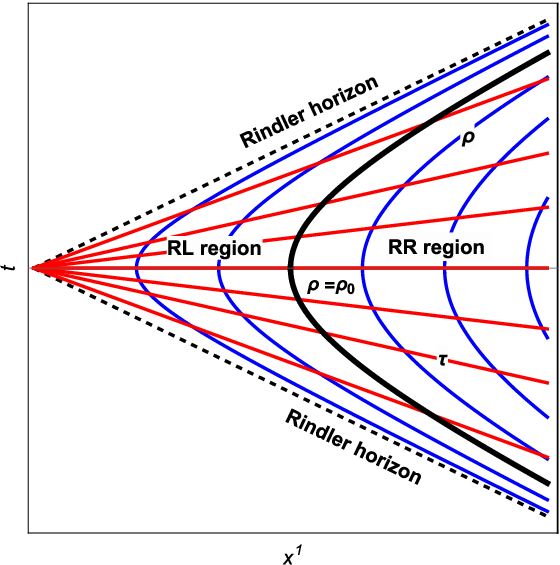,width=7.5cm,height=7.5cm}
\end{center}
\caption{The coordinate lines $\protect\tau =\mathrm{const}$ and $\protect%
\rho =\mathrm{const}$ in the R region of the Rindler spacetime on the plane $%
(x^{1},t)$. The worldline of a uniformly accelerating mirror is depicted by
a thick line. It separates the RL and RR regions.}
\label{fig1}
\end{figure}

Our main interest is the shift in the vacuum expectation values (VEVs) of
the physical characteristics of the Fulling-Rindler vacuum induced by a
uniformly accelerated mirror moving along the axis $x^{1}$ with a proper
acceleration $1/\rho _{0}$. On the boundary $\rho =\rho _{0}$ with the
normal $n_{\mu }$ the Dirac field obeys the bag boundary condition%
\begin{equation}
\left( 1+in_{\mu }\gamma ^{\mu }\right) \psi =0.  \label{BC}
\end{equation}%
In the MIT bag model of hadrons, this boundary condition is used to confine
quarks within a finite volume. It ensures that there is no flux of fermions
at the boundary on which the boundary condition is imposed. From this
condition, it follows that $n_{\mu }j^{\mu }=0$ for $\rho =\rho _{0}$, where
$j^{\mu }=\bar{\psi}\gamma ^{\mu }\psi $ being the current density. Here and
below, $\bar{\psi}=\psi ^{\dagger }\gamma ^{(0)}$ is the Dirac conjugate for
the field $\psi $. Due to the condition (\ref{BC}), the boundary $\rho =\rho
_{0}$ reflects all the fermion modes. In this sense, the boundary can be
called a mirror for the Dirac field. In the following, the terms boundary
and mirror will be used interchangeably. We assume that the mirror is
located in the R wedge of the Rindler geometry. It divides the R region into
two subregions: region $0\leq \rho <\rho _{0}$ (RL region) and region $\rho
_{0}<\rho <\infty $ (RR region). For the components of the normal in those
regions we have $n_{\mu }=\delta ^{\mathrm{(J)}}\delta _{\mu }^{1}$ with $%
\delta ^{\mathrm{(J)}}=-1$ in the RR region and $\delta ^{\mathrm{(J)}}=1$
in the RL region. In Fig.\ref{fig1} we plotted the coordinate lines in the R
wedge of the Rindler spacetime. The worldline of the mirror, separating the
RL and RR regions, is plotted by a thick curve.

As physical characteristics of the fermionic Fulling-Rindler vacuum, the
fermion condensate and the expectation value of the energy-momentum tensor
will be considered. The evaluation procedure we will employ is based on the
summation of the corresponding mode sums over the complete set of fermionic
modes. In what follows we specify the mode functions and the eigenvalues of
quantum numbers for the RL and RR regions.

\subsection{Mode functions: General structure}

We want to find a complete set of solutions to equation (\ref{eom}) for the
geometry (\ref{ds2}) obeying the boundary condition (\ref{BC}). The
expression for the spin connection is simplified to $\Gamma _{\mu }=\delta
_{\mu }^{0}\gamma ^{(0)}\gamma ^{(1)}/2$. Following the procedure described
in \cite{Bell23}, the positive and negative energy modes, $\psi _{\sigma
}^{(+)}$ and $\psi _{\sigma }^{(-)}$, respectively, are presented in the form%
\begin{equation}
\psi _{\sigma }^{(\pm )}\left( x^{\mu }\right) =N_{\sigma }e^{\mp i\omega
\tau +i\mathbf{k}\cdot \mathbf{x}}Z_{i\omega \mp \frac{1}{2}\gamma
^{(0)}\gamma ^{(1)}}\left( \lambda \rho \right) \chi _{\eta }^{(\pm )}(%
\mathbf{k}),  \label{ModePl}
\end{equation}%
where $\mathbf{k}=(k^{2},\ldots ,k^{D})$ and the function $Z_{\nu }\left(
x\right) $ is the linear combination of the modified Bessel functions $%
I_{\nu }\left( x\right) $ and $K_{\nu }\left( x\right) $:%
\begin{equation}
Z_{\nu }\left( x\right) =C_{1\sigma }I_{\nu }\left( x\right) +C_{2\sigma
}K_{\nu }\left( x\right) .  \label{ZIK}
\end{equation}%
In (\ref{ModePl}), $\sigma =(\omega ,\mathbf{k},\eta )$ presents the set of
quantum numbers specifying the modes, and
\begin{equation}
\lambda =\sqrt{k^{2}+m^{2}},\;k=|\mathbf{k}|.  \label{lam}
\end{equation}%
The quantum number $\eta $ specifies the spinorial degrees of freedom. For
the energy of the mode (\ref{ModePl}), measured by an observer with fixed
spatial coordinates $(\rho ,\mathbf{x})$, one has $\varepsilon _{\rho
}=\omega /\rho $.

The spinors $\chi _{\eta }^{(\pm )}(\mathbf{k})$ obey the relation%
\begin{equation}
P(\mathbf{k})\chi _{\eta }^{(\pm )}(\mathbf{k})=\chi _{\eta }^{(\pm )}(%
\mathbf{k}),  \label{Relxi1}
\end{equation}%
where%
\begin{equation}
P(\mathbf{k})=\frac{1}{2}\left( 1-i\gamma ^{(1)}\frac{\boldsymbol{\gamma \,}%
\mathbf{k}+m}{\lambda }\right) ,\;\boldsymbol{\gamma \,}\mathbf{k}%
=\sum_{i=2}^{D}\gamma ^{(i)}k^{i}.  \label{Pk}
\end{equation}%
For the operator (\ref{Pk}), the properties $P^{2}(\mathbf{k})=P(\mathbf{k})$
and $P^{\dagger }(\mathbf{k})=P(\mathbf{k})$ are easily checked. The
orthonormality and completeness relations%
\begin{align}
\chi _{\eta }^{(\pm )\dagger }(\mathbf{k})\chi _{\eta ^{\prime }}^{(\pm )}(%
\mathbf{k}) &=\delta _{\eta \eta ^{\prime }},  \notag \\
\sum_{\eta }\chi _{\eta \alpha }^{(\pm )}(\mathbf{k})\chi _{\eta \beta
}^{(\pm )\dagger }(\mathbf{k}) &=P_{\alpha \beta }(\mathbf{k}),
\label{RelComp}
\end{align}%
take place for $\chi _{\eta }^{(\pm )}(\mathbf{k})$, with $\alpha $ and $%
\beta $ being spinor indices. In addition, by using (\ref{Relxi1}), the
property
\begin{equation}
\left( 1+i\gamma ^{(1)}\frac{\boldsymbol{\gamma \,}\mathbf{k}+m}{\lambda }%
\right) \chi _{\eta }^{(\pm )}(\mathbf{k})=0,  \label{Rel0}
\end{equation}%
is easily proven.

The boundary condition for the modes (\ref{ModePl}) is written as%
\begin{equation}
\left( 1+i\delta ^{\mathrm{(J)}}\gamma ^{(1)}\right) Z_{i\omega \mp \frac{1}{%
2}\gamma ^{(0)}\gamma ^{(1)}}\left( \lambda \rho \right) \chi _{\eta }^{(\pm
)}(\mathbf{k})=0.  \label{BC2}
\end{equation}%
By using the relation (see also \cite{Cand78})%
\begin{equation}
f(\gamma ^{(0)}\gamma ^{(1)})=\frac{1}{2}\sum_{\varkappa =\pm 1}\left(
1+\varkappa \gamma ^{(0)}\gamma ^{(1)}\right) f(\varkappa ),  \label{rel2}
\end{equation}%
for a given function $f(x)$, the condition is rewritten as%
\begin{equation}
\sum_{\varkappa =\pm 1}\left[ 1+\varkappa \gamma ^{(0)}\gamma ^{(1)}+i\delta
^{\mathrm{(J)}}\left( \gamma ^{(1)}+\varkappa \gamma ^{(0)}\right) \right]
\chi _{\eta }^{(\pm )}(\mathbf{k})Z_{i\omega \mp \frac{\varkappa }{2}}\left(
\lambda \rho \right) =0.
\end{equation}%
Multplying by $\chi _{\eta }^{(\pm )\dagger }(\mathbf{k})$ and summing over $%
\eta $ we get%
\begin{equation}
\sum_{\varkappa =\pm 1}\sum_{\eta }\chi _{\eta }^{(+)\dagger }(\mathbf{k})%
\left[ 1+i\delta ^{\mathrm{(J)}}\left( \varkappa \gamma ^{(0)}+\gamma
^{(1)}\right) \right] \chi _{\eta }^{(\pm )}(\mathbf{k})Z_{i\omega \mp \frac{%
\varkappa }{2}}\left( \lambda \rho _{0}\right) =0,  \label{BC5}
\end{equation}%
where the relation
\begin{equation}
\chi _{\eta ^{\prime }}^{(\pm )\dagger }(\mathbf{k})\gamma ^{(0)}\gamma
^{(1)}\chi _{\eta }^{(\pm )}(\mathbf{k})=0,  \label{rel22}
\end{equation}%
has been used. This relation follows from (\ref{Rel0}). The further
simplification of (\ref{BC5}) is done by using the relations%
\begin{align}
& \sum_{\eta }\chi _{\eta }^{(j)\dagger }(\mathbf{k})\chi _{\eta }^{(j)}(%
\mathbf{k}) =\frac{N}{2},  \notag \\
& \sum_{\eta }\chi _{\eta }^{(j)\dagger }(\mathbf{k})\left( \varkappa \gamma
^{(0)}+\gamma ^{(1)}\right) \chi _{\eta }^{(j)}(\mathbf{k}) =\frac{imN}{%
2\lambda },  \label{rel23}
\end{align}%
with the result%
\begin{equation}
Z_{i\omega -\frac{1}{2}}\left( \lambda \rho _{0}\right) +Z_{i\omega +\frac{1%
}{2}}\left( \lambda \rho _{0}\right) =0,  \label{BC7}
\end{equation}%
for both positive and negative energy modes.

The mode functions (\ref{ModePl}) are normalized by the condition
\begin{equation}
\int d^{D}x\,\sqrt{|g|}\bar{\psi}_{\sigma ^{\prime }}^{(\pm )}\gamma
^{0}\psi _{\sigma }^{(\pm )}=\delta _{\sigma \sigma ^{\prime }},  \label{NC}
\end{equation}%
where $g$ is the determinant of the metric tensor. Here, $\delta _{\sigma
\sigma ^{\prime }}$ is understood as Dirac delta function for continuous
quantum numbers and as Kronecker symbol for discrete ones.

We have described the features of the modes which are the same for the RR
and RL regions. In the further consideration one needs to specify the region.

\subsection{Mode functions in the RR region}

In the RR region, the function $I_{i\omega -\frac{1}{2}\gamma ^{(0)}\gamma
^{(1)}}\left( \lambda \rho \right) $ exponentially increases for large $\rho
$ and for normalizable modes we should take $C_{1\sigma }=0$. The mode
functions are expressed as
\begin{equation}
\psi _{\sigma }^{(j)}=N_{\sigma }^{\mathrm{RR}}e^{-ji\omega \tau +i\mathbf{k}%
\cdot \mathbf{x}}K_{i\omega -j\frac{1}{2}\gamma ^{(0)}\gamma ^{(1)}}\left(
\lambda \rho \right) \chi _{\eta }^{(j)}(\mathbf{k}),  \label{ModeRR}
\end{equation}%
with $j=+,-$. For the $\rho $-dependent part we have the relations
\begin{align}
\left[ K_{i\omega \mp \frac{1}{2}\gamma ^{(0)}\gamma ^{(1)}}\left( \lambda
\rho \right) \right] ^{\dagger }& =K_{i\omega \pm \frac{1}{2}\gamma
^{(0)}\gamma ^{(1)}}\left( \lambda \rho \right) ,  \notag \\
\gamma ^{(0)}K_{i\omega \mp \frac{1}{2}\gamma ^{(0)}\gamma ^{(1)}}\left(
\lambda \rho \right) & =K_{i\omega \pm \frac{1}{2}\gamma ^{(0)}\gamma
^{(1)}}\left( \lambda \rho \right) \gamma ^{(0)}.  \label{relK}
\end{align}%
The eigenvalues of the quantum number $\omega $ are determined by the
equation (see (\ref{BC7}))
\begin{equation}
K_{i\omega -\frac{1}{2}}\left( \lambda \rho _{0}\right) +K_{i\omega +\frac{1%
}{2}}\left( \lambda \rho _{0}\right) =0.  \label{BCRR}
\end{equation}%
Note that this equation can also be written in the form $\mathrm{Re}%
\,[K_{i\omega +\frac{1}{2}}\left( \lambda \rho _{0}\right) ]=0$. We will
denote the roots of the equation (\ref{BCRR}) for a given $\lambda $ by $%
\omega =\omega _{n}=\omega _{n}(\lambda )$, $n=1,2,\ldots $. By taking into
account that $K_{i\omega _{n}\pm 1/2}\left( \lambda \rho _{0}\right) =\pm
K_{i\omega _{n}+1/2}\left( \lambda \rho _{0}\right) $, we get%
\begin{equation}
K_{i\omega _{n}\mp \frac{1}{2}\gamma ^{(0)}\gamma ^{(1)}}\left( \lambda \rho
_{0}\right) =\pm \gamma ^{(0)}\gamma ^{(1)}K_{i\omega _{n}-\frac{1}{2}%
}\left( \lambda \rho _{0}\right) .  \label{rel26}
\end{equation}

The normalization condition (\ref{NC}) for the modes (\ref{ModeRR}) is
written in the form%
\begin{equation}
J_{\eta \eta ^{\prime },nn^{\prime }}^{\mathrm{(RR)}}\left\vert N_{\sigma }^{%
\mathrm{RR}}\right\vert ^{2}=\frac{\delta _{nn^{\prime }}\delta _{\eta \eta
^{\prime }}}{\left( 2\pi \right) ^{D-1}},  \label{NC1}
\end{equation}%
where%
\begin{equation}
J_{\eta \eta ^{\prime },nn^{\prime }}^{\mathrm{(RR)}}=\chi _{\eta ^{\prime
}}^{(+)\dagger }(\mathbf{k})\int_{\rho _{0}}^{\infty }d\rho \,K_{i\omega
_{n^{\prime }}+\frac{1}{2}\gamma ^{(0)}\gamma ^{(1)}}\left( \lambda \rho
\right) K_{i\omega _{n}-\frac{1}{2}\gamma ^{(0)}\gamma ^{(1)}}\left( \lambda
\rho \right) \chi _{\eta }^{(+)}(\mathbf{k}).  \label{JRR}
\end{equation}%
The coefficient of $\left\vert N_{\sigma }^{\mathrm{RR}}\right\vert ^{2}$ is
expressed in terms of $J_{\omega \omega ^{\prime }}$, defined by (\ref{J1}),
with $\omega =\omega _{n}$, $\omega ^{\prime }=\omega _{n^{\prime }}$, $\rho
_{1}\rightarrow \rho _{0}$, $\rho _{2}=\infty $, and $Z_{\nu }(x)=K_{\nu
}(x) $. This gives%
\begin{equation}
J_{\eta \eta ^{\prime },nn^{\prime }}^{\mathrm{(RR)}}=\frac{\delta _{\eta
\eta ^{\prime }}}{2i}\lim_{\rho \rightarrow \rho _{0}}\rho \sum_{\varkappa
=\pm 1}\frac{\varkappa K_{i\omega -\varkappa /2}(\lambda \rho )K_{i\omega
^{\prime }+\varkappa /2}(\lambda \rho )}{\omega ^{\prime }-\omega }.
\label{RelNRR}
\end{equation}%
For $n^{\prime }\neq n$, we directly put $\rho =\rho _{0}$ in the arguments
of the Macdonald function and from the boundary condition it follows that,
as expected, $\sum_{\varkappa =\pm 1}\varkappa K_{i\omega _{n}-\varkappa
/2}(\lambda \rho )K_{i\omega _{n^{\prime }}+\varkappa /2}(\lambda \rho )=0$.
In the limit $\omega ^{\prime }\rightarrow \omega $, we evaluate the
right-hand side of (\ref{RelNRR}) by using the L'H\^{o}pital's rule. By
taking into account that $\varkappa K_{i\omega _{n}-\varkappa /2}(\lambda
\rho _{0})=-K_{i\omega _{n}+1/2}(\lambda \rho _{0})$, we get
\begin{equation}
J_{\eta \eta ^{\prime },nn}^{\mathrm{(RR)}}=\delta _{\eta \eta ^{\prime }}%
\frac{i}{2}\rho _{0}K_{i\omega _{n}+1/2}(\lambda \rho _{0})\partial _{\omega
}\bar{K}_{i\omega +1/2}(\lambda \rho _{0})|_{\omega =\omega _{n}},
\label{RelNRR2}
\end{equation}%
with the notation%
\begin{equation}
\bar{K}_{i\omega +1/2}(z)=K_{i\omega +1/2}(z)+K_{i\omega -1/2}(z)\text{.}
\label{NotKbar}
\end{equation}

From the Wronskian for the modified Bessel functions, by taking into account
that $K_{i\omega _{n}-1/2}\left( \lambda \rho _{0}\right) =-K_{i\omega
_{n}+1/2}\left( \lambda \rho _{0}\right) $, one finds%
\begin{equation}
K_{i\omega _{n}+\frac{1}{2}}\left( \lambda \rho _{0}\right) =-\frac{1}{%
\lambda \rho _{0}\bar{I}_{i\omega _{n}+1/2}\left( \lambda \rho _{0}\right) },
\label{KIrel}
\end{equation}%
where
\begin{equation}
\bar{I}_{i\omega +1/2}(z)=I_{i\omega +1/2}(z)-I_{i\omega -1/2}(z).
\label{NotIbar}
\end{equation}%
The final expression for $J_{\eta \eta ^{\prime },nn^{\prime }}^{\mathrm{(RR)%
}}$ reads%
\begin{equation}
J_{\eta \eta ^{\prime },nn^{\prime }}^{\mathrm{(RR)}}=\delta _{\eta \eta
^{\prime }}\delta _{nn^{\prime }}\frac{\partial _{\omega }\bar{K}_{i\omega
+1/2}(\lambda \rho _{0})}{2i\lambda \bar{I}_{i\omega +1/2}\left( \lambda
\rho _{0}\right) }|_{\omega =\omega _{n}}.  \label{JRR2}
\end{equation}%
With this result, the normalization coefficient is obtained from (\ref{NC1}):%
\begin{equation}
\left\vert N_{\sigma }^{\mathrm{RR}}\right\vert ^{2}=\frac{2i\lambda \bar{I}%
_{i\omega +1/2}\left( \lambda \rho _{0}\right) }{\left( 2\pi \right)
^{D-1}\partial _{\omega }\bar{K}_{i\omega +1/2}\left( \lambda \rho
_{0}\right) }|_{\omega =\omega _{n}}.  \label{NRR}
\end{equation}%
Note that $\bar{K}_{-i\omega +1/2}(z)=\bar{K}_{i\omega +1/2}(z)$ and, hence,
the function $\bar{K}_{i\omega +1/2}(z)$ is real for real values of $\omega $
and $z$. In addition, we have $K_{1/2-i\omega _{n}}\left( \lambda \rho
_{0}\right) =-K_{1/2+i\omega _{n}}\left( \lambda \rho _{0}\right) $ and the
function $K_{1/2+i\omega _{n}}\left( \lambda \rho _{0}\right) $ is purely
imaginary. From (\ref{KIrel}) it follows that the same is the case for the
function $\bar{I}_{i\omega _{n}+1/2}\left( \lambda \rho _{0}\right) $ and
the function $i\bar{I}_{i\omega _{n}+1/2}\left( \lambda \rho _{0}\right) $
in (\ref{NRR}) is real.

\subsection{Mode functions in the RL region}

In the RL region, corresponding to $0<\rho <\rho _{0}$, both the functions $%
I_{\nu }\left( x\right) $ and $K_{\nu }\left( x\right) $ are allowed in (\ref%
{ZIK}). The relative coefficient in the linear combination of those
functions is determined by the boundary condition (\ref{BC7}). The mode
functions are presented in the form%
\begin{equation}
\psi _{\sigma }^{(j)}=N_{\sigma }^{\mathrm{RL}}e^{-ji\omega \tau +i\mathbf{k}%
\cdot \mathbf{x}}Z_{i\omega -j\frac{1}{2}\gamma ^{(0)}\gamma ^{(1)}}\left(
\lambda \rho \right) \chi _{\eta }^{(j)}(\mathbf{k}),  \label{ModeRL}
\end{equation}%
where%
\begin{equation}
Z_{\nu }\left( \lambda \rho \right) =e^{-\nu \pi i}I_{\nu }\left( \lambda
\rho \right) +C_{\sigma }K_{\nu }\left( \lambda \rho \right) .  \label{ZnuRL}
\end{equation}%
Note that the function $Z_{\nu }\left( \lambda \rho \right) $ obeys the same
recurrence relations as the function $K_{\nu }\left( \lambda \rho \right) $
(the factor $e^{-\nu \pi i}$ in front of $I_{\nu }\left( \lambda \rho
\right) $ is added for this purpose). The coefficient $C_{\sigma }$ is found
from the boundary condition (\ref{BC7}) and is given by the expression%
\begin{equation}
C_{\sigma }=ie^{\pi \omega }\frac{\bar{I}_{i\omega +1/2}(\lambda \rho _{0})}{%
\bar{K}_{i\omega +1/2}(\lambda \rho _{0})},  \label{CRL}
\end{equation}%
with the notations (\ref{NotKbar}) and (\ref{NotIbar}). Note that unlike the
RR region, the spectrum of $\omega $ in the RL region is continuous. It can
be shown that
\begin{equation}
C_{\sigma }^{\ast }=C_{\sigma }+\frac{i}{\pi }\left( e^{2\pi \omega
}+1\right) ,  \label{Crel}
\end{equation}%
where the star stands for the complex conjugate.

By using the relation (\ref{Crel}) we can see that (compare with (\ref{relK}%
))%
\begin{align}
\left[ Z_{i\omega \mp \frac{1}{2}\gamma ^{(0)}\gamma ^{(1)}}\left( \lambda
\rho \right) \right] ^{\dagger }& =Z_{i\omega \pm \frac{1}{2}\gamma
^{(0)}\gamma ^{(1)}}\left( \lambda \rho \right) ,  \notag \\
\gamma ^{(0)}Z_{i\omega \mp \frac{1}{2}\gamma ^{(0)}\gamma ^{(1)}}\left(
\lambda \rho \right) & =Z_{i\omega \pm \frac{1}{2}\gamma ^{(0)}\gamma
^{(1)}}\left( \lambda \rho \right) \gamma ^{(0)},  \label{ZRL}
\end{align}%
and the normalization condition for the positive energy modes (\ref{ModeRL})
is written in the form%
\begin{equation}
J_{\eta \eta ^{\prime },\omega \omega ^{\prime }}^{\mathrm{(RL)}}\left\vert
N_{\sigma }^{\mathrm{RL}}\right\vert ^{2}=\frac{\delta \left( \omega -\omega
^{\prime }\right) }{\left( 2\pi \right) ^{D-1}}\delta _{\eta \eta ^{\prime
}},  \label{NCRL}
\end{equation}%
where
\begin{equation}
J_{\eta \eta ^{\prime },\omega \omega ^{\prime }}^{\mathrm{(RL)}}=\chi
_{\eta ^{\prime }}^{(+)\dagger }(\mathbf{k})\int_{0}^{\rho _{0}}d\rho
\,Z_{i\omega ^{\prime }+\frac{1}{2}\gamma ^{(0)}\gamma ^{(1)}}\left( \lambda
\rho \right) Z_{i\omega -\frac{1}{2}\gamma ^{(0)}\gamma ^{(1)}}\left(
\lambda \rho \right) \chi _{\eta }^{(+)}(\mathbf{k}).  \label{JRL}
\end{equation}%
By using the result (\ref{J14}) with $\rho _{2}\rightarrow \rho _{0}$, $\rho
_{1}\rightarrow 0$, and the boundary condition (\ref{BC7}), we can see that
the contribution from the upper limit $\rho =\rho _{0}$ becomes zero. The
expression (\ref{JRL}) is transformed to
\begin{equation}
J_{\eta \eta ^{\prime },\omega \omega ^{\prime }}^{\mathrm{(RL)}}=\frac{%
\delta _{\eta \eta ^{\prime }}}{2i\lambda }\lim_{x\rightarrow
0}x\sum_{\varkappa =\pm 1}\frac{\varkappa Z_{i\omega -\varkappa
/2}(x)Z_{i\omega ^{\prime }+\varkappa /2}(x)}{\omega ^{\prime }-\omega },
\label{JRL2}
\end{equation}%
where we have introduced $x=\lambda \rho $. By making use of the asymptotics
for the modified Bessel functions for small values of the argument and the
relation (\ref{Crel}), we can see that for $0<x\ll 1$, to the leading order,
\begin{equation}
\sum_{\varkappa =\pm 1}\varkappa Z_{i\omega -\frac{\varkappa }{2}%
}(x)Z_{i\omega ^{\prime }+\frac{\varkappa }{2}}(x)\approx i\pi \left\vert
C_{\sigma }\right\vert ^{2}\frac{\sin \left[ \left( \omega -\omega ^{\prime
}\right) \ln (x/2)\right] }{x\cosh \left( \omega \pi \right) }.
\label{ZZrel}
\end{equation}%
Plugging this in (\ref{JRL2}) and by taking into account that
\begin{equation}
\lim_{x\rightarrow 0}\frac{\sin \left( u\ln (x/2)\right) }{u}=-\pi \delta
\left( u\right) ,  \label{delt}
\end{equation}%
we get%
\begin{equation}
J_{\eta \eta ^{\prime },\omega \omega ^{\prime }}^{\mathrm{(RL)}}=\frac{%
\delta _{\eta \eta ^{\prime }}\pi ^{2}\left\vert C_{\sigma }\right\vert ^{2}%
}{2\lambda \cosh \left( \omega \pi \right) }\delta \left( \omega ^{\prime
}-\omega \right) .  \label{JRL3}
\end{equation}%
With this result, the normalization constant in the RL region is found to be%
\begin{equation}
\left\vert N_{\sigma }^{\mathrm{RL}}\right\vert ^{2}=\frac{8\lambda \cosh
\left( \omega \pi \right) }{\left( 2\pi \right) ^{D+1}\left\vert C_{\sigma
}\right\vert ^{2}},  \label{NRL}
\end{equation}%
with $C_{\sigma }$ given by (\ref{CRL}).

\section{Fermion condensate}

\label{sec:FC}

After specifying the normalized mode functions in the RR and RL regions, we
investigate the VEVs of physical observables. In this section, we consider
the fermion condensate. The formation of a nonzero fermion condensate is
critically important in quantum field theory and in many condensed matter
systems exhibiting superconductivity and superfluidity. The condensate
provides a mechanism that gives fermions mass and acts as an order parameter
for dynamical symmetry breaking, such as chiral symmetry breaking in quantum
chromodynamics through the formation of a quark condensate. In the context
of the Nambu-Jona-Lasinio model with quartic interaction, the effective
dynamical mass of fermions is proportional to the fermion condensate, and
the coefficient is determined by the four-fermion coupling constant.

The fermion condensate is defined as the VEV $\left\langle 0\right\vert \bar{%
\psi}\psi \left\vert 0\right\rangle \equiv \left\langle \bar{\psi}\psi
\right\rangle $, where $\left\vert 0\right\rangle $ stands for the
Fulling-Rindler vacuum state. Expanding the field operator in terms of
complete set of modes $\{\psi _{\sigma }^{(+)},\psi _{\sigma }^{(-)}\}$, the
condensate is presented in the form of the mode-sum
\begin{equation}
\left\langle \bar{\psi}\psi \right\rangle =\frac{1}{2}\sum_{\sigma }\left(
\bar{\psi}_{\sigma }^{(-)}\psi _{\sigma }^{(-)}-\bar{\psi}_{\sigma
}^{(+)}\psi _{\sigma }^{(+)}\right) ,  \label{FCn}
\end{equation}%
where $\sum_{\sigma }$ includes a summation over discrete quantum numbers
and an integration over the continuous ones. The fermion condensate in the
RR and RL regions will be studied separately.

\subsection{Condensate in the RR region}

Plugging the mode functions (\ref{ModeRR}) in (\ref{FCn}) and using the
second relation in (\ref{rel23}), the fermion condensate in the RR-region is
presented in the form%
\begin{equation}
\left\langle \bar{\psi}\psi \right\rangle =-\frac{mN}{2^{D}\pi ^{D-1}}\int d%
\mathbf{k}\sum_{n=1}^{\infty }\frac{\bar{I}_{i\omega +1/2}\left( \lambda
\rho _{0}\right) }{\partial _{\omega }\bar{K}_{i\omega +1/2}\left( \lambda
\rho _{0}\right) }\left[ K_{i\omega +\frac{1}{2}}^{2}\left( \lambda \rho
\right) -K_{i\omega -\frac{1}{2}}^{2}\left( \lambda \rho \right) \right]
_{\omega =\omega _{n}}.  \label{FCRR}
\end{equation}%
This expression is divergent and some regularization scheme is implicitly
assumed (e.g., point-splitting or a cutoff function). For points $\rho \neq
\rho _{0}$, the divergences are the same as those in a problem without
boundaries and the renormalization of the condensate reduces to the
corresponding procedure in the problem without boundaries. This can also be
understood based on general considerations. In quantum field theories in
curved spaces, the divergences of vacuum averages at a given spacetime point
are uniquely determined by the local geometric characteristics of the
background spacetime at that point, such as the Riemann tensor and
combinations constructed from it. Introducing boundaries does not change the
local geometry of spacetime at points outside the boundary; therefore, it
does not lead to new divergences in the local physical characteristics of
vacuum averages. Thus, renormalizing these averages is equivalent to
renormalizing them in a problem without boundaries. In the problem under
consideration, the background spacetime is flat, and subtracting the
Minkowski spacetime parts from the vacuum expectation values of local
observables yields finite physical values at points outside the boundary.

The above arguments do not apply to points on the boundary. Typically, local
vacuum expectation values diverge as they approach the boundary. In the
literature on the Casimir effect, these divergences are studied for various
boundary and background geometries (see, for example, \cite{Most97}-\cite%
{Casi11}). The presence of surface divergences indicates that physical
quantities containing contributions from local averages at boundary points
necessitate additional renormalization. An important example of such a
quantity is the total vacuum energy. In the following discussion we will
explicitly separate the boundary-free contributions in local observables
such as the fermion condensate and the mean energy-momentum tensor. The
corresponding renormalization procedure has already been discussed in \cite%
{Bell23}. Thus, the choice of a specific regularization scheme is not
important for the subsequent analysis.

From the point of view of further calculations, the representation (\ref%
{FCRR}) has two disadvantages: (i) the eigenvalues $\omega _{n}$ of the
energy are given implicitly, as roots of the equation (\ref{BCRR}), and (ii)
the series terms for large $n$ are highly oscillatory. Both of these
difficulties can be overcome by applying to the series over $n$ in (\ref%
{FCRR}) the summation formula (\ref{Sum}) with the function%
\begin{equation}
F(z)=K_{iz+\frac{1}{2}}^{2}\left( \lambda \rho \right) -K_{iz-\frac{1}{2}%
}^{2}\left( \lambda \rho \right) .  \label{FzRR}
\end{equation}%
This allows to present the fermion condensate in the form%
\begin{equation}
\left\langle \bar{\psi}\psi \right\rangle =\left\langle \bar{\psi}\psi
\right\rangle _{0}+\frac{mN}{\left( 2\pi \right) ^{D}}\int d\mathbf{k}%
\int_{0}^{\infty }dx\,\frac{\bar{I}_{x+1/2}(\lambda \rho _{0})}{\bar{K}%
_{x+1/2}(\lambda \rho _{0})}\left[ K_{x+\frac{1}{2}}^{2}\left( \lambda \rho
\right) -K_{x-\frac{1}{2}}^{2}\left( \lambda \rho \right) \right] ,
\label{FCRR2}
\end{equation}%
where%
\begin{equation}
\left\langle \bar{\psi}\psi \right\rangle _{0}=\frac{4mN}{\left( 2\pi
\right) ^{D+1}}\int d\mathbf{k}\int_{0}^{\infty }d\omega \,\cosh \pi \omega
\,\mathrm{Im}\left[ K_{\frac{1}{2}-i\omega }^{2}\left( \lambda \rho \right) %
\right] .  \label{FC0}
\end{equation}%
The part $\left\langle \bar{\psi}\psi \right\rangle _{0}$ comes from the
integral in the square brackets of (\ref{Sum}) and coincides with the
fermion condensate in the boundary-free geometry. The representation (\ref%
{FCRR2}) is well-adapted for investigating the properties of the condensate.
First, explicit knowledge of the eigenmodes $\omega _{n}$ is not required.
Second, the integrand monotonically decreases within the range of the
integration variables near the upper limits of integration. Third, the
boundary-induced contribution is explicitly separated.

The renormalization of (\ref{FC0}) is done by subtracting the condensate in
Minkowski spacetime: $\left\langle \bar{\psi}\psi \right\rangle _{0}^{%
\mathrm{ren}}=\left\langle \bar{\psi}\psi \right\rangle _{0}-\left\langle
\bar{\psi}\psi \right\rangle _{\mathrm{M}}$ and the renormalized fermion
condensate for the Fulling-Rindler vacuum in the absence of boundaries is
expressed as \cite{Bell23}%
\begin{equation}
\left\langle \bar{\psi}\psi \right\rangle _{0}^{\mathrm{ren}}=\frac{2^{1-D}mN%
}{\pi ^{\frac{D+3}{2}}\Gamma \left( \frac{D-1}{2}\right) }\int_{0}^{\infty
}d\omega \,e^{-\pi \omega }\int_{m}^{\infty }d\lambda \mathbf{\,}\lambda
\left( \lambda ^{2}-m^{2}\right) ^{\frac{D-3}{2}}\,\mathrm{Im}\left[
K_{1/2-i\omega }^{2}\left( \lambda \rho \right) \right] ,  \label{FC0r}
\end{equation}%
for $D\geq 2$. For $D=1$ we have
\begin{equation}
\left\langle \bar{\psi}\psi \right\rangle _{0}^{\mathrm{ren}}=\frac{m}{\pi
^{2}}\int_{0}^{\infty }d\omega \,e^{-\pi \omega }\,\mathrm{Im}\left[
K_{1/2-i\omega }^{2}\left( m\rho \right) \right] .  \label{FC0rD1}
\end{equation}%
Another representation for the boundary-free part in the fermion condensate
is provided in \cite{Bell23}:%
\begin{equation}
\left\langle \bar{\psi}\psi \right\rangle _{0}^{\mathrm{ren}}=-\frac{2Nm^{D}%
}{\left( 2\pi \right) ^{\frac{D+3}{2}}}\int_{0}^{\infty }du\,\frac{u\sinh u}{%
u^{2}+\pi ^{2}/4}f_{\frac{D-1}{2}}\left( 2m\rho \cosh u\right) ,
\label{FC0r2}
\end{equation}%
with the function
\begin{equation}
f_{\nu }\left( z\right) =z^{-\nu }K_{\nu }\left( z\right) .  \label{fnu1}
\end{equation}%
The expression (\ref{FC0r2}) shows that for a massive field the
boundary-free part in the fermion condensate is negative.

Integrating the boundary-induced contribution in (\ref{FCRR2}) over the
angular coordinates of the momentum $\mathbf{k}$, the renormalized fermion
condensate in the RR region is decomposed as%
\begin{equation}
\left\langle \bar{\psi}\psi \right\rangle _{\mathrm{ren}}=\left\langle \bar{%
\psi}\psi \right\rangle _{0}^{\mathrm{ren}}+\left\langle \bar{\psi}\psi
\right\rangle _{\mathrm{b}},  \label{FCdec}
\end{equation}%
where the boundary-induced contribution is given by the formula
\begin{align}
\left\langle \bar{\psi}\psi \right\rangle _{\mathrm{b}}& =\frac{2^{1-D}mN}{%
\pi ^{\frac{D+1}{2}}\Gamma \left( \frac{D-1}{2}\right) }\int_{m}^{\infty
}d\lambda \,\lambda \left( \lambda ^{2}-m^{2}\right) ^{\frac{D-3}{2}}  \notag
\\
& \times \int_{1/2}^{\infty }d\nu \,\frac{\bar{I}_{\nu }(\lambda \rho _{0})}{%
\bar{K}_{\nu }(\lambda \rho _{0})}\left[ K_{\nu }^{2}\left( \lambda \rho
\right) -K_{\nu -1}^{2}\left( \lambda \rho \right) \right] ,  \label{FCRR3}
\end{align}%
for $D\geq 2$. In the case $D=1$, the boundary-induced part is expressed as
\begin{equation}
\left\langle \bar{\psi}\psi \right\rangle _{\mathrm{b}}=\frac{m}{\pi }%
\int_{1/2}^{\infty }d\nu \,\frac{\bar{I}_{\nu }(m\rho _{0})}{\bar{K}_{\nu
}(m\rho _{0})}\left[ K_{\nu }^{2}\left( m\rho \right) -K_{\nu -1}^{2}\left(
m\rho \right) \right] .  \label{FCRRD1}
\end{equation}%
In the integration range of (\ref{FCRR3}) we have the relations%
\begin{equation}
\bar{I}_{\nu }(z)<0,\;\bar{K}_{\nu }(z)>0,\;K_{\nu }\left( z\right) >K_{\nu
-1}\left( z\right) .  \label{RelsIK}
\end{equation}%
We conclude from this that, in the RR region, both the boundary-free and
boundary-induced contributions to the fermion condensate are negative for a
massive field. In the zero-mass limit, the condensate tends to zero like $%
\left\langle \bar{\psi}\psi \right\rangle _{\mathrm{b}}\propto m$ for $D>2$
and like $\left\langle \bar{\psi}\psi \right\rangle _{\mathrm{b}}\propto
m\ln m$ for $D=2$. In the case $D=1$, the condensate tends to nonzero
limiting value. Taking the limit $m\rightarrow 0$ in (\ref{FCRRD1}), we
obtain%
\begin{equation}
\left\langle \bar{\psi}\psi \right\rangle _{\mathrm{b}}|_{m=0}=-\frac{1}{%
2\pi \rho \ln \left( \rho /\rho _{0}\right) },\;D=1.  \label{FCD1m0}
\end{equation}

Let us consider the asymptotics of the fermion condensate near the plate and
at large distances from it. Near the boundary, assuming that $\rho /\rho
_{0}-1\ll 1$, the dominant contribution to (\ref{FCRR3}) comes from large
values of $\lambda $ and $\nu $. Introducing a new integration variable $%
z=\lambda /\nu $, we use the uniform asymptotic expansions of the modified
Bessel functions for large values of the order (see, for example, \cite%
{Abra72}). It can be seen that for $\nu \gg 1$, to the leading order,
\begin{align}
\frac{\bar{I}_{\nu }(\nu u)}{\bar{K}_{\nu }(\nu u)} &\approx \frac{e^{2\nu
\eta (u)}}{\pi u}\left( u-t_{u}\right) \left( t_{u}+1\right) ,  \notag \\
K_{\nu }^{2}(\nu u)-K_{\nu -1}^{2}(\nu u) &\approx \frac{\pi e^{-2\nu \eta
(u)}}{\nu u^{2}}\left( 1-\frac{1}{t_{u}}\right)  \label{IKas}
\end{align}%
where
\begin{equation}
\eta (u)=t_{u}+\ln \frac{u}{1+t_{u}},\;t_{u}=\sqrt{1+u^{2}}.  \label{etau}
\end{equation}%
Substituting these asymptotics in (\ref{FCRR3}) and by taking into account
that for $\rho /\rho _{0}-1\ll 1$ one has $\eta \left( u\rho /\rho
_{0}\right) -\eta \left( u\right) \approx t_{u}\left( \rho /\rho
_{0}-1\right) $, for the leading term in the expansion of the fermion
condensate over the distance from the boundary we get%
\begin{equation}
\left\langle \bar{\psi}\psi \right\rangle _{\mathrm{b}}\approx \frac{\left(
4\pi \right) ^{-\frac{D+1}{2}}mN}{\left( \rho -\rho _{0}\right) ^{D-1}}\left[
\Gamma \left( \frac{D-1}{2}\right) -\frac{\Gamma \left( \frac{D}{2}-1\right)
}{\Gamma \left( \frac{D-1}{2}\right) }\Gamma \left( \frac{D}{2}\right) %
\right] ,  \label{FCnear}
\end{equation}%
for $D\geq 3$. For $D=2$ the asymptotic is given by%
\begin{equation}
\left\langle \bar{\psi}\psi \right\rangle _{\mathrm{b}}\approx \frac{\sqrt{m}%
}{4\pi \left( \rho -\rho _{0}\right) ^{3/2}}\left[ \frac{\Gamma \left( \frac{%
3}{4}\right) }{\Gamma \left( \frac{1}{4}\right) }-\frac{\Gamma \left( \frac{5%
}{4}\right) }{\Gamma \left( \frac{3}{4}\right) }\right] .  \label{FCnearD2}
\end{equation}%
In the case $D=1$ one obtains%
\begin{equation}
\left\langle \bar{\psi}\psi \right\rangle _{\mathrm{b}}\approx -\frac{1}{%
2\pi \left( \rho -\rho _{0}\right) }.  \label{FCnearD1}
\end{equation}%
This leading term does not depend on the mass and could also been obtained
directly from (\ref{FCD1m0}).

At large distances from the boundary and for a massive field, $m\rho \gg 1$,
the argument $\lambda \rho $ of the Macdonald functions in (\ref{FCRRD1}) is
large and we use the corresponding asymptotic \cite{Abra72}. The dominant
contribution to the integral over $\lambda $ comes from the region near the
lower limit of the integral. In the leading order for $D\geq 1$ one obtains%
\begin{equation}
\left\langle \bar{\psi}\psi \right\rangle _{\mathrm{b}}\approx \frac{%
Nm^{D}e^{-2m\rho }}{2^{D+1}\pi ^{\frac{D-1}{2}}\left( m\rho \right) ^{\frac{%
D+3}{2}}}\int_{1/2}^{\infty }d\nu \,\left( 2\nu -1\right) \frac{\bar{I}_{\nu
}(m\rho _{0})}{\bar{K}_{\nu }(m\rho _{0})}.  \label{FClarge}
\end{equation}%
The asymptotic of the boundary-free part in the limit $m\rho \gg 1$ reads%
\begin{equation}
\left\langle \bar{\psi}\psi \right\rangle _{0}^{\mathrm{ren}}\approx -\frac{%
Nm^{D}e^{-2m\rho }}{2^{D+1}\pi ^{\frac{D+5}{2}}\left( m\rho \right) ^{\frac{%
D+3}{2}}}.  \label{FC0large}
\end{equation}%
For the ratio of the boundary-induced and boundary-free contributions we get%
\begin{equation}
\frac{\left\langle \bar{\psi}\psi \right\rangle _{\mathrm{b}}}{\left\langle
\bar{\psi}\psi \right\rangle _{0}^{\mathrm{ren}}}\approx -\pi
^{3}\int_{1/2}^{\infty }d\nu \,\left( 2\nu -1\right) \frac{\bar{I}_{\nu
}(m\rho _{0})}{\bar{K}_{\nu }(m\rho _{0})}.  \label{FCrel}
\end{equation}%
This ratio does not depend on the number of spatial dimensions. The
numerical evaluation shows that the right-hand side in (\ref{FCrel}) is
smaller than 1 for $m\rho _{0}<0.04$ and larger than 1 for $m\rho _{0}>0.04$%
. For $m\rho _{0}=1$ the ratio (\ref{FCrel}) is near 40 and it increases
with increasing $m\rho _{0}$. Hence, in the range $m\rho _{0}>0.04$, the
total VEV at large distances is dominated by the boundary-induced part. For
a massless field the fermion condensate vanishes in spatial dimensions $%
D\geq 2$ and the large-distance asymptotic for $D=1$ is given by (\ref%
{FCD1m0}).

\subsection{Condensate in the RL region}

With the mode functions (\ref{ModeRL}), the fermion condensate in the
RL-region is expressed by the formula
\begin{equation}
\left\langle \bar{\psi}\psi \right\rangle =\frac{2imN}{\left( 2\pi \right)
^{D+1}}\int_{0}^{\infty }d\omega \int d\mathbf{k\,}\frac{\cosh \left( \omega
\pi \right) }{\left\vert C_{\sigma }\right\vert ^{2}}\left[ Z_{i\omega +%
\frac{1}{2}}^{2}\left( \lambda \rho \right) -Z_{i\omega -\frac{1}{2}%
}^{2}\left( \lambda \rho \right) \right] ,  \label{FCLR}
\end{equation}%
where $C_{\sigma }$ is defined in (\ref{CRL}) and%
\begin{equation}
Z_{i\omega \pm \frac{1}{2}}\left( \lambda \rho \right) =C_{\sigma
}K_{i\omega \pm \frac{1}{2}}\left( \lambda \rho \right) \mp e^{\pi \omega
}iI_{i\omega \pm \frac{1}{2}}\left( \lambda \rho \right) .  \label{ZRL2}
\end{equation}%
It is easy to see that%
\begin{equation}
Z_{i\omega \pm \frac{1}{2}}\left( \lambda \rho \right) =\left[ Z_{i\omega
\mp \frac{1}{2}}\left( \lambda \rho \right) \right] ^{\ast },  \label{ZRL3}
\end{equation}%
and the expression in the right-hand side of (\ref{FCLR}) is explicitly
real. In order to separate the boundary-free contribution we use the identity%
\begin{equation}
\sum_{\kappa =\pm 1}\kappa \frac{Z_{i\omega +\frac{\kappa }{2}}^{2}\left(
\lambda \rho \right) }{\left\vert C_{\sigma }\right\vert ^{2}}=\sum_{\kappa
=\pm 1}\kappa \left[ K_{i\omega +\frac{\kappa }{2}}^{2}\left( \lambda \rho
\right) +\frac{\pi \bar{K}_{i\omega +\frac{1}{2}}(\lambda \rho _{0})}{2\cosh
\left( \omega \pi \right) }\sum_{j=\pm 1}\frac{jI_{ji\omega +\frac{\kappa }{2%
}}^{2}\left( \lambda \rho \right) }{\bar{I}_{ji\omega +\frac{1}{2}}(\lambda
\rho _{0})}\right] .  \label{Ident}
\end{equation}%
The proof of this relation is presented in Appendix \ref{sec:AppIdent}. The
identity (\ref{Ident}) allows to write the fermion condensate (\ref{FCLR})
in the form%
\begin{equation}
\left\langle \bar{\psi}\psi \right\rangle =\left\langle \bar{\psi}\psi
\right\rangle _{0}+\frac{imN}{2\left( 2\pi \right) ^{D}}\int d\mathbf{k}%
\sum_{\kappa ,j=\pm 1}\kappa j\int_{0}^{\infty }d\omega \frac{\bar{K}%
_{ji\omega +\frac{1}{2}}(\lambda \rho _{0})}{\bar{I}_{ji\omega +\frac{1}{2}%
}(\lambda \rho _{0})}I_{ji\omega +\frac{\kappa }{2}}^{2}\left( \lambda \rho
\right) ,  \label{FCLR1}
\end{equation}%
with the boundary-free condensate $\left\langle \bar{\psi}\psi \right\rangle
_{0}$ given by (\ref{FC0}).

For the further transformation of the boundary-induced part in (\ref{FCLR1}%
), we rotate the integration contour over $\omega $ by the angle $\pi /2$
for the term $j=-1$ and by the angle $-\pi /2$ for the term $j=1$. For the
integrals we get%
\begin{equation}
\int_{0}^{\infty }d\omega \frac{\bar{K}_{ji\omega +\frac{1}{2}}(\lambda \rho
_{0})}{\bar{I}_{ji\omega +\frac{1}{2}}(\lambda \rho _{0})}I_{ji\omega +\frac{%
\kappa }{2}}^{2}\left( \lambda \rho \right) =-ji\int_{0}^{\infty }dx\frac{%
\bar{K}_{x+\frac{1}{2}}(\lambda \rho _{0})}{\bar{I}_{x+\frac{1}{2}}(\lambda
\rho _{0})}I_{x+\frac{\kappa }{2}}^{2}\left( \lambda \rho \right) ,
\label{Ints}
\end{equation}%
and the fermion condensate is presented as%
\begin{equation}
\left\langle \bar{\psi}\psi \right\rangle =\left\langle \bar{\psi}\psi
\right\rangle _{0}+\frac{mN}{\left( 2\pi \right) ^{D}}\int d\mathbf{k}%
\int_{0}^{\infty }dx\frac{\bar{K}_{x+1/2}(\lambda \rho _{0})}{\bar{I}%
_{x+1/2}(\lambda \rho _{0})}\left[ I_{x+\frac{1}{2}}^{2}\left( \lambda \rho
\right) -I_{x-\frac{1}{2}}^{2}\left( \lambda \rho \right) \right] .
\label{FCLR2}
\end{equation}%
For $D\geq 2$, integrating over the angular part of the momentum, the
condensate is decomposed as (\ref{FCdec}), where the boundary-induced
contribution is given by the expression%
\begin{align}
\left\langle \bar{\psi}\psi \right\rangle _{\mathrm{b}}& =\frac{2^{1-D}mN}{%
\pi ^{\frac{D+1}{2}}\Gamma \left( \frac{D-1}{2}\right) }\int_{m}^{\infty
}d\lambda \,\lambda \left( \lambda ^{2}-m^{2}\right) ^{\frac{D-3}{2}}  \notag
\\
& \times \int_{1/2}^{\infty }d\nu \,\frac{\bar{K}_{\nu }(\lambda \rho _{0})}{%
\bar{I}_{\nu }(\lambda \rho _{0})}\left[ I_{\nu }^{2}\left( \lambda \rho
\right) -I_{\nu -1}^{2}\left( \lambda \rho \right) \right] .  \label{FCLR3}
\end{align}%
The renormalized boundary-free condensate is expressed by the formula (\ref%
{FC0r}). In the case $D=1$ we get%
\begin{equation}
\left\langle \bar{\psi}\psi \right\rangle _{\mathrm{b}}=\frac{m}{\pi }%
\int_{1/2}^{\infty }d\nu \frac{\bar{K}_{\nu }(m\rho _{0})}{\bar{I}_{\nu
}(m\rho _{0})}\left[ I_{\nu }^{2}\left( m\rho \right) -I_{\nu -1}^{2}\left(
m\rho \right) \right] .  \label{FCRLD1}
\end{equation}%
By taking into account the relations (\ref{RelsIK}) and $I_{\nu }\left(
z\right) <I_{\nu -1}\left( z\right) $, it is seen that the boundary-induced
contribution in the fermion condensate is positive in the RL region. For a
massless field the fermion condensate vanishes in spatial dimensions $D\geq
2 $ and is given by
\begin{equation}
\left\langle \bar{\psi}\psi \right\rangle _{\mathrm{b}}=\frac{1}{2\pi \rho
\ln (\rho _{0}/\rho )},  \label{FCRLD1m0}
\end{equation}%
for $D=1$. Comparing with (\ref{FCD1m0}), we see that in this special case
the expressions for the condensates in the RL and RR regions differ by the
sign.

The asymptotic near the boundary is found in the way similar to that we have
used for the RR-region, by using the uniform asymptotic expansions of the
modified Bessel functions. By taking into account (\ref{IKas}) and
\begin{equation}
I_{\nu }^{2}(\nu u)-I_{\nu -1}^{2}(\nu u)\approx -\frac{e^{2\nu \eta (u)}}{%
\pi \nu u^{2}}\left( 1+\frac{1}{t_{u}}\right) ,  \label{Inuas}
\end{equation}%
we can show that in the leading order
\begin{equation}
\left\langle \bar{\psi}\psi \right\rangle _{\mathrm{b}}\approx \frac{\left(
4\pi \right) ^{-\frac{D+1}{2}}mN}{\left( \rho _{0}-\rho \right) ^{D-1}}\left[
\Gamma \left( \frac{D-1}{2}\right) +\frac{\Gamma \left( \frac{D}{2}-1\right)
}{\Gamma \left( \frac{D-1}{2}\right) }\Gamma \left( \frac{D}{2}\right) %
\right] ,  \label{FCRLnear}
\end{equation}%
for $D\geq 3$. In the case $D=2$ one has%
\begin{equation}
\left\langle \bar{\psi}\psi \right\rangle _{\mathrm{b}}\approx \frac{\sqrt{m}%
}{4\pi \left( \rho _{0}-\rho \right) ^{3/2}}\,\left[ \frac{\Gamma \left(
\frac{5}{4}\right) }{\Gamma \left( \frac{3}{4}\right) }+\frac{\Gamma \left(
\frac{3}{4}\right) }{\Gamma \left( \frac{1}{4}\right) }\right] ,
\label{FCRLnearD2}
\end{equation}%
and for $D=1$ we get%
\begin{equation}
\left\langle \bar{\psi}\psi \right\rangle _{\mathrm{b}}\approx \frac{1}{2\pi
(\rho _{0}-\rho )}.  \label{FCRLnearD1}
\end{equation}

\begin{figure}[tbph]
\begin{center}
\begin{tabular}{cc}
\epsfig{figure=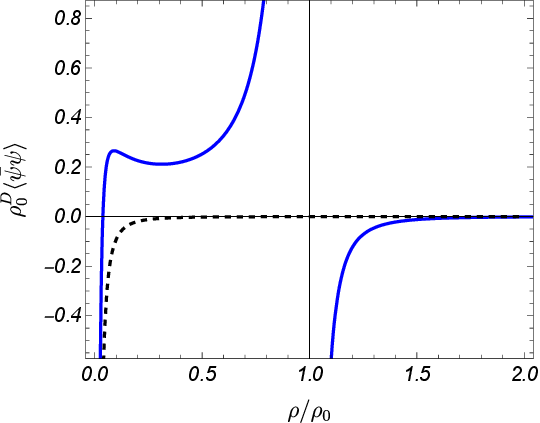,width=8cm,height=6.5cm} & \quad %
\epsfig{figure=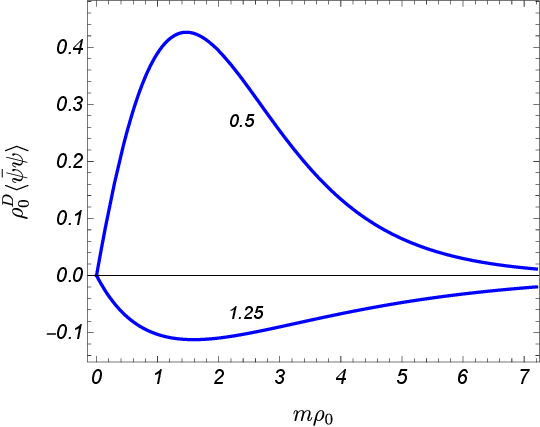,width=8cm,height=6.5cm}%
\end{tabular}%
\end{center}
\caption{The fermion condensate for spatial dimension $D=3$ in the RR and RL
regions versus $\protect\rho /\protect\rho _{0}$ (left panel) and $m\protect%
\rho _{0}$ (right panel). The graphs on the left panel are plotted for $m%
\protect\rho _{0}=0.5$ and the dashed curve presents the condensate in the
boundary-free geometry. The numbers near the curves on the right panel
present the values of $\protect\rho /\protect\rho _{0}$.}
\label{fig2}
\end{figure}

In Fig. \ref{fig2}, we present the total fermion condensate (\ref{FCdec}),
in units of $1/\rho _{0}^{D}$, as a function of the ratio $\rho /\rho _{0}$
(left panel) and of the mass (right panel) for spatial dimension $D=3$. On
the left panel, the graphs for the RR and RL regions are plotted for $m\rho
_{0}=0.5$ and the dashed curve corresponds to the fermion condensate in the
boundary-free problem (given by (\ref{FC0r2})). The numbers near the curves
on the right panel correspond to the values of the ratio $\rho /\rho _{0}$.
As already concluded based on asymptotic analysis, the fermion condensate is
dominated by the boundary-induced part near the mirror and by the
boundary-free contribution near the Rindler horizon.

From the asymptotic analysis presented above, it follows that for the taken
value of the combination $m\rho _{0}$, the contribution imposed by the
boundary dominates over the entire range $\rho _{0}<\rho <\infty $. The
boundary-free contribution dominates only in the region close to the
horizon. An important feature of the fermion condensate in the RL region is
that it changes sign at a certain value of acceleration. In models with
interacting fields, changing the sign of the fermion condensate can have
important consequences. They include tachyonic instabilities and spontaneous
symmetry breaking in models with interacting scalar and fermionic fields and
phase transition in Nambu-Jona-Lasinio-type models with quartic fermionic
interactions.

\section{VEV of the energy-momentum tensor}

\label{sec:EMT}

In this section, we will consider another important local characteristic of
the vacuum state: the VEV of the energy-momentum tensor, $\left\langle
0\right\vert T_{\mu \alpha }\left\vert 0\right\rangle \equiv \left\langle
T_{\mu \alpha }\right\rangle $. In addition to describing the distribution
of the vacuum energy density and stresses, it acts as a source of the
gravitational field in the semiclassical Einstein equations. The
corresponding mode-sum formula reads
\begin{equation}
\langle T_{\mu \alpha }\rangle =-\frac{i}{4}\sum_{\sigma }\sum_{j=+,-}j[\bar{%
\psi}_{\sigma }^{(j)}(x)\gamma _{(\mu }\nabla _{\alpha )}\psi _{\sigma
}^{(j)}(x)-(\nabla _{(\mu }\bar{\psi}_{\sigma }^{(j)}(x))\gamma _{\alpha
)}\psi _{\sigma }^{(j)}(x)]\ ,  \label{EMT}
\end{equation}%
where the brackets denote symmetrization over the enclosed indices and%
\begin{equation}
\nabla _{\mu }\bar{\psi}_{\sigma }^{(j)}=\partial _{\mu }\bar{\psi}_{\sigma
}^{(j)}-\bar{\psi}_{\sigma }^{(j)}\Gamma _{\mu }.  \label{Nabl}
\end{equation}%
In both RR and RL regions the vacuum energy-momentum tensor is diagonal.
From the problem symmetry we also expect that
\begin{equation}
\langle T_{2}^{2}\rangle =\langle T_{3}^{3}\rangle =\cdots =\langle
T_{D}^{D}\rangle .  \label{Stress}
\end{equation}%
Of course, these relations are checked by direct evaluation.

\subsection{RR region}

By using the mode functions (\ref{ModeRR}) and the relations for the spinors
$\chi _{\eta }^{(j)}(\mathbf{k})$ given above, the diagonal components of
the vacuum energy-momentum tensor in the RR region are presented in the form
(no summation over $\mu $)%
\begin{equation}
\langle T_{\mu }^{\mu }\rangle =-\frac{N}{2^{D}\pi ^{D-1}}\int d\mathbf{k}%
\sum_{n=1}^{\infty }\frac{\bar{I}_{i\omega +\frac{1}{2}}\left( \lambda \rho
_{0}\right) }{\partial _{\omega }\bar{K}_{i\omega +\frac{1}{2}}\left(
\lambda \rho _{0}\right) }F_{\mu }[K_{i\omega +\frac{1}{2}}\left( \lambda
\rho \right) ]|_{\omega =\omega _{n}}.  \label{EMTRR}
\end{equation}%
Here, for a given function $G_{\nu }(z)$, the notations%
\begin{align}
F_{0}\left[ G_{\nu }\left( \lambda \rho \right) \right] & =\delta _{G}\frac{%
\lambda }{\rho }\left( 2\nu -1\right) G_{\nu }\left( \lambda \rho \right)
G_{\nu -1}\left( \lambda \rho \right) ,  \notag \\
F_{1}\left[ G_{\nu }\left( \lambda \rho \right) \right] & =\delta
_{G}\lambda ^{2}\left[ G_{\nu }^{\prime }\left( \lambda \rho \right) G_{\nu
-1}\left( \lambda \rho \right) -G_{\nu }\left( \lambda \rho \right) G_{\nu
-1}^{\prime }\left( \lambda \rho \right) \right] ,  \label{FRR1} \\
F_{l}\left[ G_{\nu }\left( \lambda \rho \right) \right] & =\frac{\lambda
^{2}-m^{2}}{D-1}\left[ G_{\nu -1}^{2}\left( \lambda \rho \right) -G_{\nu
}^{2}\left( \lambda \rho \right) \right] ,  \notag
\end{align}%
with $l=2,3,\ldots ,D$, are introduced and the prime stands for the
derivative with respect to the argument of the function. In the discussion
below, as a function $G_{\nu }\left( z\right) $ we will take the functions $%
K_{\nu }\left( z\right) $, $Z_{\nu }\left( z\right) $, and $I_{\nu }\left(
z\right) $. For those functions we define%
\begin{equation}
\delta _{G}=\left\{
\begin{array}{ll}
1, & G=K,Z \\
-1, & G=I%
\end{array}%
\right. .  \label{delG}
\end{equation}%
The function $F_{l,i\omega }[G_{x+1/2}\left( \lambda \rho \right) ]$ for the
components $\langle T_{l}^{l}\rangle $, with $l=2,3,\ldots ,D$, obtained by
the direct evaluation from (\ref{EMT}) has the form $\left( k^{l}\right)
^{2}[G_{x-1/2}^{2}\left( \lambda \rho \right) -G_{x+1/2}^{2}\left( \lambda
\rho \right) ]$. We have replaced $\left( k^{l}\right) ^{2}$ by $\left(
\lambda ^{2}-m^{2}\right) /(D-1)$ by taking into account the relations (\ref%
{Stress}). In deriving the expression for those components, we have employed
the relation%
\begin{equation}
\sum_{\eta }\chi _{\eta }^{(j)\dagger }\left( \gamma ^{(0)}\pm \gamma
^{(1)}\right) \gamma ^{(l)}\chi _{\eta }^{(j)}(\mathbf{k})=\pm \frac{iN}{%
2\lambda }k^{l}.  \label{relkl}
\end{equation}%
By using the recurrence formulas for the modified Bessel functions, it can
be seen that the relation%
\begin{equation}
F_{1}\left[ G_{\nu }\left( \lambda \rho \right) \right] =\frac{1-D}{%
1-m^{2}/\lambda ^{2}}F_{2}\left[ G_{\nu }\left( \lambda \rho \right) \right]
-F_{0}\left[ G_{\nu }\left( \lambda \rho \right) \right]  \label{RelFi}
\end{equation}%
holds between the functions (\ref{FRR1}).

For the series over $n$ in (\ref{EMTRR}) we use the summation formula (\ref%
{Sum}). The part in the VEV coming from the integral in the square brackets
of (\ref{Sum}) corresponds to the VEV of the energy-momentum tensor in the
boundary-free problem. Denoting it by $\langle T_{\mu }^{\mu }\rangle _{0}$,
the VEV is decomposed as (no summation over $\mu $)%
\begin{equation}
\langle T_{\mu }^{\mu }\rangle =\langle T_{\mu }^{\mu }\rangle _{0}+N\int
\frac{d\mathbf{k}}{\left( 2\pi \right) ^{D}}\int_{0}^{\infty }dx\,\frac{\bar{%
I}_{x+\frac{1}{2}}(\lambda \rho _{0})}{\bar{K}_{x+\frac{1}{2}}(\lambda \rho
_{0})}F_{\mu }[K_{x+\frac{1}{2}}\left( \lambda \rho \right) ],
\label{EMTRR2}
\end{equation}%
with
\begin{equation}
\langle T_{\mu }^{\mu }\rangle _{0}=\frac{2iN}{\left( 2\pi \right) ^{D+1}}%
\int d\mathbf{k}\int_{0}^{\infty }d\omega \,\cosh \left( \pi \omega \right)
F_{\mu }[K_{i\omega +\frac{1}{2}}\left( \lambda \rho \right) ].  \label{EMT0}
\end{equation}%
For points away from the boundary, the renormalization is reduced to the
renormalization of (\ref{EMT0}). Similar to the fermion condensate, that is
done by subtracting the VEV of the energy-momentum tensor for the Minkowski
vacuum: $\langle T_{\mu }^{\alpha }\rangle _{0}^{\mathrm{ren}}=\langle
T_{\mu }^{\alpha }\rangle _{0}-\langle T_{\mu }^{\alpha }\rangle _{\mathrm{M}%
}$. The renormalized VEV takes the form (no summation over $\mu $) \cite%
{Bell23}
\begin{equation}
\langle T_{\mu }^{\mu }\rangle _{0}^{\mathrm{ren}}=\frac{2^{-D}iN}{\pi ^{%
\frac{D+3}{2}}\Gamma \left( \frac{D-1}{2}\right) }\int_{0}^{\infty }d\omega
\,e^{-\pi \omega }\int_{m}^{\infty }d\lambda \,\mathbf{\,}\lambda \left(
\lambda ^{2}-m^{2}\right) ^{\frac{D-3}{2}}F_{\mu }[K_{i\omega +\frac{1}{2}%
}\left( \lambda \rho \right) ],  \label{Tmu2}
\end{equation}%
for $D\geq 2$. After integration over the angular coordinates of $\mathbf{k}$
in (\ref{EMTRR2}), the renormalized VEV in the RR region is presented in the
decomposed form (no summation over $\mu $)%
\begin{equation}
\langle T_{\mu }^{\mu }\rangle _{\mathrm{ren}}=\langle T_{\mu }^{\mu
}\rangle _{0}^{\mathrm{ren}}+\langle T_{\mu }^{\mu }\rangle _{\mathrm{b}},
\label{EMTdec}
\end{equation}%
with the boundary-induced part%
\begin{equation}
\langle T_{\mu }^{\mu }\rangle _{\mathrm{b}}=\frac{2^{1-D}N}{\pi ^{\frac{D+1%
}{2}}\Gamma \left( \frac{D-1}{2}\right) }\int_{m}^{\infty }d\lambda
\,\lambda \left( \lambda ^{2}-m^{2}\right) ^{\frac{D-3}{2}%
}\int_{1/2}^{\infty }d\nu \,\frac{\bar{I}_{\nu }(\lambda \rho _{0})}{\bar{K}%
_{\nu }(\lambda \rho _{0})}F_{\mu }[K_{\nu }\left( \lambda \rho \right) ].
\label{EMTRR3}
\end{equation}%
In the special case $D=1$ one has (no summation over $\mu $)%
\begin{equation}
\langle T_{\mu }^{\mu }\rangle _{\mathrm{b}}=\frac{1}{\pi }%
\int_{1/2}^{\infty }d\nu \,\frac{\bar{I}_{\nu }(m\rho _{0})}{\bar{K}_{\nu
}(m\rho _{0})}F_{\mu }[K_{\nu }\left( m\rho \right) ],  \label{EMTRRD1}
\end{equation}%
and%
\begin{equation}
\langle T_{\mu }^{\mu }\rangle _{0}^{\mathrm{ren}}=\frac{i}{2\pi ^{2}}%
\int_{0}^{\infty }d\omega \,e^{-\pi \omega }\,F_{\mu }[K_{i\omega +\frac{1}{2%
}}\left( m\rho \right) ],  \label{EMTRR0D1}
\end{equation}%
with $\mu =0,1$, and in the expressions (\ref{FRR1}) $\lambda $ is replaced
by $m$. Simpler representation for the boundary-free VEV is given in \cite%
{Bell23} (no summation over $l$):%
\begin{align}
\langle T_{0}^{0}\rangle _{0}^{\mathrm{ren}}& =-\frac{2Nm^{D+1}}{\left( 2\pi
\right) ^{\frac{D+3}{2}}}\int_{0}^{\infty }du\,\frac{u\sinh u}{u^{2}+\pi
^{2}/4}\,\left[ f_{\frac{D-1}{2}}(2m\rho \cosh u)+Df_{\frac{D+1}{2}}(2m\rho
\cosh u)\right] ,  \notag \\
\langle T_{l}^{l}\rangle _{0}^{\mathrm{ren}}& =\frac{2Nm^{D+1}}{\left( 2\pi
\right) ^{\frac{D+3}{2}}}\int_{0}^{\infty }du\,\frac{u\sinh u}{u^{2}+\pi
^{2}/4}f_{\frac{D+1}{2}}\left( 2m\rho \cosh u\right) ,  \label{Tmu0}
\end{align}%
with $l=1,2,\ldots ,D$, and the functioin $f_{\nu }(x)$ is defined by (\ref%
{fnu1}). This shows that the vacuum stresses are isotropic in the
boundary-free problem. The vacuum effective pressure along the $l$th
direction is given by $p_{l}^{(0)}=-\langle T_{l}^{l}\rangle _{0}^{\mathrm{%
ren}}$. As seen from (\ref{Tmu0}), in the boundary-free problem both the
energy density and the pressure in the Fulling-Rindler vacuum are negative.

We can check that the VEV of the energy-momentum tensor obeys the continuity
equation $\nabla _{\alpha }\langle T_{\mu }^{\alpha }\rangle _{\mathrm{ren}%
}=0$. For the problem under consideration it is reduced to the relation $%
\langle T_{0}^{0}\rangle _{\mathrm{ren}}=\partial _{\rho }\left( \rho
\langle T_{1}^{1}\rangle _{\mathrm{ren}}\right) $. By using the differential
equation for the function $K_{\nu }(z)$ it is easily seen that $\partial
_{\rho }\left( \rho F_{1}[K_{\nu }\left( \lambda \rho \right) ]\right)
=F_{0}[K_{\nu }\left( \lambda \rho \right) ]$ and, hence, the continuity
equation indeed takes place. In addition, for the functions $F_{\mu
}[K_{x+1/2}\left( \lambda \rho \right) ]$ we have%
\begin{equation}
\sum_{\mu =0}^{D}F_{\mu }[K_{x+\frac{1}{2}}\left( \lambda \rho \right)
]=m^{2}\left[ K_{x+\frac{1}{2}}^{2}\left( \lambda \rho \right) -K_{x-\frac{1%
}{2}}^{2}\left( \lambda \rho \right) \right] .  \label{TraceF}
\end{equation}%
From here it follows that the VEV of the energy-momentum tensor obeys the
trace relation%
\begin{equation}
\langle T_{\mu }^{\mu }\rangle _{\mathrm{ren}}=m\left\langle \bar{\psi}\psi
\right\rangle _{\mathrm{ren}}.  \label{Trace}
\end{equation}%
Comparing the expressions for the fermion condensate and the parallel
stresses, we can check that the following relation takes place (no summation
over $l=2,3,\ldots ,D$)%
\begin{equation}
\langle T_{l}^{l}\rangle _{0}^{\mathrm{ren}}=-\frac{\pi }{m}\left\langle
\bar{\psi}\psi \right\rangle _{(D+3),0}^{\mathrm{ren}},\;\langle
T_{l}^{l}\rangle _{\mathrm{b}}=-\frac{\pi }{m}\left\langle \bar{\psi}\psi
\right\rangle _{(D+3),\mathrm{b}},  \label{relFC}
\end{equation}%
where $\left\langle \bar{\psi}\psi \right\rangle _{(D+3),0}^{\mathrm{ren}}$
and $\left\langle \bar{\psi}\psi \right\rangle _{(D+3),\mathrm{b}}$ are the
fermion condensates in $(D+3)$-dimensional Rindler spacetime, given by (\ref%
{FC0r2}) and (\ref{Tmu0}) with the replacements $D\rightarrow D+2$ (note
that $N_{D+3}=2N_{D+1}$).

For a massless fermionic field the vacuum energy-momentum tensor is
traceless. In this special case the boundary-induced energy-momentum tensor
is simplified to (no summation over $\mu $)%
\begin{equation}
\langle T_{\mu }^{\mu }\rangle _{\mathrm{b}}=\frac{N\left( 2\rho \right)
^{1-D}}{\pi ^{\frac{D+1}{2}}\Gamma \left( \frac{D-1}{2}\right) }%
\int_{0}^{\infty }du\,u^{D-2}\int_{1/2}^{\infty }d\nu \,\frac{\bar{I}_{\nu
}(u\rho _{0}/\rho )}{\bar{K}_{\nu }(u\rho _{0}/\rho )}F_{\mu }[K_{\nu
}\left( u\right) ],  \label{EMTRRm0}
\end{equation}%
where the functions $F_{\mu }[K_{\nu }\left( u\right) ]$ are given by (\ref%
{FRR1}) with $G=K$, $m=0$, and $\lambda =u/\rho $. In (\ref{EMTRRm0}), an
additional factor $1/\rho ^{2}$ comes from the functions $F_{\mu }[K_{\nu
}\left( u\right) ]$ and the combination $\rho ^{D+1}\langle T_{\mu }^{\mu
}\rangle _{\mathrm{b}}$ is a function of only the ratio $\rho _{0}/\rho $.
The boundary-free contribution is expressed as \cite{Bell23} (no summation
over $l$)%
\begin{equation}
\langle T_{l}^{l}\rangle _{0}^{\mathrm{ren}}=-\frac{1}{D}\langle
T_{0}^{0}\rangle _{0}^{\mathrm{ren}}=\frac{2N\Gamma \left( \frac{D+1}{2}%
\right) }{\left( 4\pi \right) ^{\frac{D+3}{2}}\rho ^{D+1}}\int_{0}^{\infty
}du\,\frac{u\sinh u}{\left( u^{2}+\frac{\pi ^{2}}{4}\right) \cosh ^{D+1}u}\,,
\label{Tmu0m0}
\end{equation}%
with $l=1,2,\ldots ,D$.

From the relations (\ref{RelsIK}) it follows that the boundary-induced
contribution in the vacuum energy density is negative and the corresponding
stresses $\langle T_{l}^{l}\rangle _{\mathrm{b}}$, $l=2,3,\ldots ,D$, are
positive in the RR region:%
\begin{equation}
\langle T_{0}^{0}\rangle _{\mathrm{b}}<0,\;\langle T_{2}^{2}\rangle _{%
\mathrm{b}}>0,\;\rho >\rho _{0}.  \label{T00sign}
\end{equation}%
Note that the same inequalities take place for the boundary-free parts (\ref%
{Tmu0}).

Now, let us consider the asymptotic behavior of the components of the
energy-momentum tensor in the RR region. In the way similar to that we have
used for the fermion condensate, it can be shown that for points near the
boundary, $\rho /\rho _{0}-1\ll 1$, the leading term of the boundary-induced
mean energy density is expressed as
\begin{equation}
\langle T_{0}^{0}\rangle _{\mathrm{b}}\approx \frac{\left( D-1\right) N}{%
2\left( 4\pi \right) ^{\frac{D+1}{2}}\left( \rho -\rho _{0}\right) ^{D+1}}%
\left[ \Gamma \left( \frac{D+1}{2}\right) -\frac{\Gamma \left( \frac{D}{2}%
\right) }{\Gamma \left( \frac{D+1}{2}\right) }\Gamma \left( \frac{D}{2}%
+1\right) \right] .  \label{T00near}
\end{equation}%
For the normal and parallel stresses one gets (no summation over $%
l=2,3,\ldots ,D$)%
\begin{equation}
\langle T_{1}^{1}\rangle _{\mathrm{b}}\approx -\frac{\rho /\rho _{0}-1}{D}%
\langle T_{0}^{0}\rangle _{\mathrm{b}},\;\langle T_{l}^{l}\rangle _{\mathrm{b%
}}\approx -\frac{\langle T_{0}^{0}\rangle _{\mathrm{b}}}{D-1}.
\label{T11near}
\end{equation}%
As seen, near the boundary we have $|\langle T_{1}^{1}\rangle _{\mathrm{b}%
}|\ll |\langle T_{0}^{0}\rangle _{\mathrm{b}}|$. In the region under
consideration the total VEV is dominated by the boundary-induced part.

The large-distance asymptotic of the boundary-induced energy-momentum tensor
for a massive field, assuming $m\rho \gg 1$, is given by (no summation over $%
l=1,2,\ldots ,D$)
\begin{equation}
\langle T_{0}^{0}\rangle _{\mathrm{b}}\approx -2m\rho \langle
T_{l}^{l}\rangle _{\mathrm{b}}\approx \frac{Nm^{D+1}e^{-2m\rho }}{2^{D+1}\pi
^{\frac{D-1}{2}}\left( m\rho \right) ^{\frac{D+3}{2}}}\int_{1/2}^{\infty
}d\nu \,\left( 2\nu -1\right) \frac{\bar{I}_{\nu }(m\rho _{0})}{\bar{K}_{\nu
}(m\rho _{0})},  \label{T00far}
\end{equation}%
with an exponential decay. Note that one has $\langle T_{0}^{0}\rangle _{%
\mathrm{b}}\approx m\left\langle \bar{\psi}\psi \right\rangle _{\mathrm{b}}$%
. The behavior of the boundary-free contribution in the same limit is
described by%
\begin{equation}
\langle T_{l}^{l}\rangle _{0}^{\mathrm{ren}}\approx -\frac{\langle
T_{0}^{0}\rangle _{0}^{\mathrm{ren}}}{2m\rho }\approx \frac{%
Nm^{D+1}e^{-2m\rho }}{2^{D+2}\left( \pi m\rho \right) ^{\frac{D+5}{2}}}.
\label{T00far0}
\end{equation}%
Both the contributions to the energy density are suppressed by the same
factor $\left( m\rho \right) ^{-\frac{D+3}{2}}e^{-2m\rho }$. In the leading
order, the vacuum stresses are isotropic and, compared to the energy
density, they contain an additional suppression factor $1/(m\rho )$. The
relative contributions of the boundary-induced and boundary-free parts in
the energy density for the region $m\rho \gg 1$ is given by the right-hand
side of (\ref{FCrel}). In the region $m\rho _{0}>0.04$ and for $m\rho \gg 1$%
, boundary-induced part dominates in the total VEV of the energy-momentum
tensor. For a massless field and $\rho \gg \rho _{0}$, the leading terms of
the asymptotic expansion are expressed as (no summation over $l=1,2,\ldots
,D $)%
\begin{equation}
\langle T_{l}^{l}\rangle _{\mathrm{b}}\approx -\frac{1}{D}\langle
T_{0}^{0}\rangle _{\mathrm{b}}\approx \frac{N\Gamma \left( \frac{D}{2}%
\right) \ln ^{-2}\left( 2\rho /\rho _{0}\right) }{2^{D+1}\pi ^{\frac{D}{2}%
+1}D\rho ^{D+1}}.  \label{T00m0far}
\end{equation}%
The boundary-free part for a massless field is given by (\ref{Tmu0m0}) and
it decays like $1/\rho ^{D+1}$. This shows that for a massless field the
boundary-free part dominates at large distances.

\subsection{RL region}

By the evaluation similar to that for the RR region, the VEV of the
energy-momentum tensor in the RL region is presented in the form (no
summation over $\mu $)
\begin{equation}
\langle T_{\mu }^{\mu }\rangle =\frac{2iN}{\left( 2\pi \right) ^{D+1}}\int d%
\mathbf{k\,}\int_{0}^{\infty }d\omega \,\frac{\cosh \left( \omega \pi
\right) }{\left\vert C_{\sigma }\right\vert ^{2}}F_{\mu }[Z_{i\omega +\frac{1%
}{2}}\left( \lambda \rho \right) ],  \label{EMTRL}
\end{equation}%
where the functions $F_{\mu }[Z_{\nu }\left( \lambda \rho \right) ]$ are
defined by (\ref{FRR1}) with the replacement $G\rightarrow Z$. The
boundary-induced contribution in the VEV (\ref{EMTRL}) is obtained by
subtracting the boundary free part $\langle T_{\mu }^{\mu }\rangle _{0}$,
given by (\ref{EMT0}). The VEV is expressed as (no summation over $\mu $)%
\begin{equation}
\langle T_{\mu }^{\mu }\rangle =\langle T_{\mu }^{\mu }\rangle _{0}+\frac{2iN%
}{\left( 2\pi \right) ^{D+1}}\int d\mathbf{k}\int_{0}^{\infty }d\omega \cosh
\left( \omega \pi \right) \left\{ \frac{F_{\mu }[Z_{i\omega +\frac{1}{2}%
}\left( \lambda \rho \right) ]}{\left\vert C_{\sigma }\right\vert ^{2}}%
-F_{\mu }[K_{i\omega +\frac{1}{2}}\left( \lambda \rho \right) ]\right\} .
\label{EMTRL2}
\end{equation}%
By using the identities given in Appendix \ref{sec:AppIdent}, for the
difference in the integrand we have%
\begin{equation}
\frac{F_{\mu }[Z_{\nu }\left( \lambda \rho \right) ]}{\left\vert C_{\sigma
}\right\vert ^{2}}-F_{\mu }[K_{\nu }\left( \lambda \rho \right)
]=\sum_{j=\pm 1}\frac{j\pi \bar{K}_{\nu }(\lambda \rho _{0})F_{\mu }\left[
I_{ji\omega +\frac{1}{2}}\left( \lambda \rho \right) \right] }{2\cosh \left(
\omega \pi \right) \bar{I}_{ji\omega +1/2}(\lambda \rho _{0})},
\label{DifRL}
\end{equation}%
where $\nu =i\omega +1/2$. Substituting this in (\ref{EMTRL2}), we rotate
the integration contour over $\omega $ in the complex $\omega $-plane by the
angle $-\pi /2$ for the term with $j=+1$ and by the angle $\pi /2$ for the
part with $j=-1$. This leads to the following represention (no summation
over $\mu $):%
\begin{equation}
\langle T_{\mu }^{\mu }\rangle =\langle T_{\mu }^{\mu }\rangle _{0}+\frac{N}{%
\left( 2\pi \right) ^{D}}\int d\mathbf{k}\int_{0}^{\infty }dx\frac{\bar{K}%
_{x+1/2}(\lambda \rho _{0})}{\bar{I}_{x+1/2}(\lambda \rho _{0})}F_{\mu }%
\left[ I_{x+\frac{1}{2}}\left( \lambda \rho \right) \right] .  \label{EMTRL3}
\end{equation}

The renormalized VEV is obtained by subtracting the VEV for the Minkowski
vacuum. The subtraction is reduced to the renormalization of the
boundary-free part and for $D\geq 2$ we get the decomposition (\ref{EMTdec})
with the boundary-induced contribution (no summation over $\mu $)%
\begin{equation}
\langle T_{\mu }^{\mu }\rangle _{\mathrm{b}}=\frac{2^{1-D}N}{\pi ^{\frac{D+1%
}{2}}\Gamma \left( \frac{D-1}{2}\right) }\int_{m}^{\infty }d\lambda
\,\lambda \left( \lambda ^{2}-m^{2}\right) ^{\frac{D-3}{2}%
}\int_{1/2}^{\infty }d\nu \frac{\bar{K}_{\nu }(\lambda \rho _{0})}{\bar{I}%
_{\nu }(\lambda \rho _{0})}F_{\mu }\left[ I_{\nu }\left( \lambda \rho
\right) \right] ,  \label{EMTRL4}
\end{equation}%
where the functions $F_{\mu }[I_{\nu }\left( \lambda \rho \right) ]$ are
defined in (\ref{FRR1}) with $G=I$ and with $\delta _{I}=-1$. In the special
case $D=1$ we get%
\begin{equation}
\langle T_{\mu }^{\mu }\rangle _{\mathrm{b}}=\frac{1}{\pi }%
\int_{1/2}^{\infty }d\nu \,\frac{\bar{I}_{\nu }(m\rho _{0})}{\bar{K}_{\nu
}(m\rho _{0})}F_{\mu }[I_{\nu }\left( m\rho \right) ].  \label{EMTRLD1}
\end{equation}%
From the relations (\ref{RelsIK}) we see that the boundary-induced energy
density is positive and the stresses $\langle T_{l}^{l}\rangle _{\mathrm{b}}$%
, $l=2,3,\ldots ,D$, are negative in the RL region:%
\begin{equation}
\langle T_{0}^{0}\rangle _{\mathrm{b}}>0,\;\langle T_{2}^{2}\rangle _{%
\mathrm{b}}<0,\;\rho <\rho _{0}.  \label{T00signRL}
\end{equation}%
These signs are opposite to those for the boundary-free contributions.
Similar to the RR region, we have the relations (\ref{relFC}) between the
normal stress in $(D+1)$-dimensional spacetime and the fermion condensate in
$(D+3)$-dimensional spacetime. The expression (\ref{EMTRL4}) is further
simplified for a massless field (no summation over $\mu $):%
\begin{equation}
\langle T_{\mu }^{\mu }\rangle _{\mathrm{b}}=\frac{N\left( 2\rho \right)
^{1-D}}{\pi ^{\frac{D+1}{2}}\Gamma \left( \frac{D-1}{2}\right) }%
\int_{0}^{\infty }du\,u^{D-2}\int_{1/2}^{\infty }d\nu \,\frac{\bar{K}_{\nu
}(u\rho _{0}/\rho )}{\bar{I}_{\nu }(u\rho _{0}/\rho )}F_{\mu }\left[ I_{\nu
}\left( u\right) \right] ,  \label{EMTRLm0}
\end{equation}%
with the function $F_{\mu }\left[ I_{\nu }\left( u\right) \right] $ from (%
\ref{RelsIK}), where $G=I$ and $\lambda =u/\rho $.

\begin{figure}[tbph]
\begin{center}
\begin{tabular}{cc}
\epsfig{figure=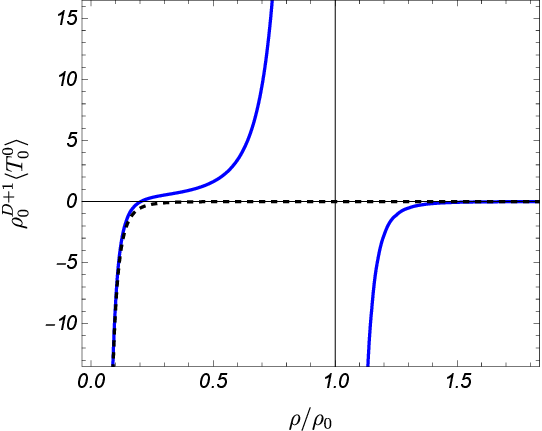,width=8.cm,height=6.5cm} & \quad %
\epsfig{figure=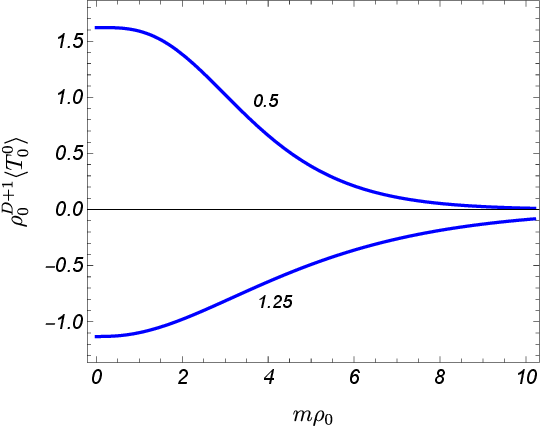,width=8.cm,height=6.5cm}%
\end{tabular}%
\end{center}
\caption{The energy density in the Fulling-Rindler vacuum for the $D=3$
Dirac field as a function of $\protect\rho /\protect\rho _{0}$ (left panel)
and $m\protect\rho _{0}$ (right panel). For the graphs on the left panel we
have taken $m\protect\rho _{0}=0.5$ and the numbers near the curves on the
right panel correspond to the values of $\protect\rho /\protect\rho _{0}$.
The dashed curve on the left panel presents the energy density in the
boundary-free problem.}
\label{fig3}
\end{figure}

Near the boundary we use the uniform asymptotic expansions for the modified
Bessel functions. The calculations similar to those for the RR region lead
to the result
\begin{equation}
\langle T_{0}^{0}\rangle _{\mathrm{b}}\approx \frac{\left( 4\pi \right) ^{-%
\frac{D+1}{2}}\left( D-1\right) N}{2\left( \rho _{0}-\rho \right) ^{D+1}}%
\left[ \Gamma \left( \frac{D+1}{2}\right) +\frac{\Gamma \left( \frac{D}{2}%
\right) }{\Gamma \left( \frac{D+1}{2}\right) }\Gamma \left( \frac{D}{2}%
+1\right) \right] ,  \label{T00RLnear}
\end{equation}%
for the boundary-induced energy density. The asymptotics for other
components are related to the energy density by the formulas (\ref{T11near}%
). Figure \ref{fig3} displays the dependence of the vacuum energy density
(in units of $1/\rho _{0}^{D+1}$) on the ratio $\rho /\rho _{0}$ (left
panel) and on $m\rho _{0}$ (right panel). The graphs are plotted for the
spatial dimension $D=3$ and for $m\rho _{0}=0.5$ on the left panel. The
dashed curve on the left panel corresponds to the vacuum energy density in
the geometry without boundaries. The numbers near the curves on the right
panel correspond to the values of the ratio $\rho /\rho _{0}$. In accordance
with the asymptotic analysis given above, near the horizon the vacuum energy
density is dominated by the boundary-free contribution and is negative. In
the region near the boundary, the energy density is dominated by the
boundary-induced contribution and it is positive in the RL region and
negative in the RR region.

\begin{figure}[tbph]
\begin{center}
\begin{tabular}{cc}
\epsfig{figure=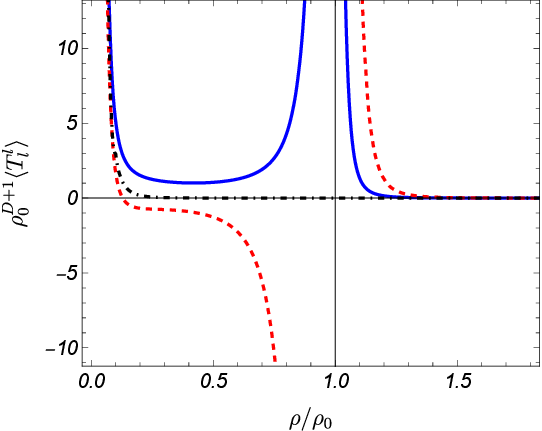,width=8.cm,height=6.5cm} & \quad %
\epsfig{figure=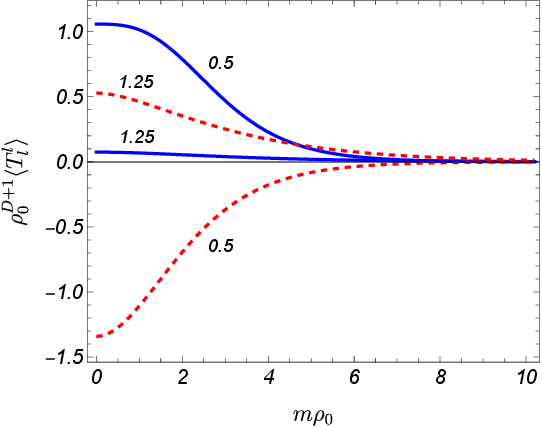,width=8.cm,height=6.5cm}%
\end{tabular}%
\end{center}
\caption{The same as in Fig. \protect\ref{fig3} for the stresses parallel
and perpendicular to the boundary (full and dashed curves, respectively).
The dot-dashed curve on the left panel corresponds to the stress in the
boundary-free problem.}
\label{fig4}
\end{figure}

In Fig. \ref{fig4} we have plotted the vacuum stresses along the directions
perpendicular and parallel to the boundary as functions of $\rho /\rho _{0}$
(left panel) and $m\rho _{0}$ (right panel). The full and dashed curves
correspond to the stresses $\langle T_{1}^{1}\rangle $ and $\langle
T_{2}^{2}\rangle $ in the model with $D=3$. For the graphs on the left panel
$m\rho _{0}=0.5$ and the numbers near the curves on the right panel present
the values of the ratio $\rho /\rho _{0}$. The vacuum stresses in the
boundary-free problem are presented by the dot-dashed curve (recall that in
the boundary-free geometry $\langle T_{1}^{1}\rangle =\langle
T_{2}^{2}\rangle $).

It is of interest to compare the results for the Fulling-Rindler vacuum with
the vacuum expectation values for a planar mirror, with bag boundary
condition, at rest relative to an inertial observer in the Minkowski vacuum.
The corresponding VEVs were investigated in Ref. \cite{Eliz11} for a more
general geometry involving two parallel plates and with toroidal
compactification of a part of spatial dimensions (see also \cite%
{Bell09b,Bell13} for the Casimir energy and the vacuum current density).
The results for a single boundary in Minkowski spacetime are obtained as a
special case. In this case, of course, the VEVs are symmetric with respect
to the plate. The background line element reads $ds_{\mathrm{M}}^{2}=\eta
_{\mu \nu }dx^{\prime \mu }dx^{\prime \nu }$, with $-\infty <x^{\prime \mu
}<+\infty $, and assuming that the boundary is located at $x^{\prime 1}=0$,
for the fermion condensate from \cite{Eliz11} we obtain%
\begin{equation}
\left\langle \bar{\psi}\psi \right\rangle _{\mathrm{ren}}^{\mathrm{(M)}}=%
\frac{Nm^{D}}{\left( 2\pi \right) ^{\frac{D+1}{2}}}\left[ f_{\frac{D-1}{2}%
}\left( 2m|x^{\prime 1}|\right) -2m|x^{\prime 1}|f_{\frac{D+1}{2}}\left(
2m|x^{\prime 1}|\right) \right] ,  \label{FCM}
\end{equation}%
where $f_{\nu }(x)$ is defined by (\ref{fnu1}). The condensate given by (\ref%
{FCM}) is negative. For the VEV of the energy-momentum tensor, after
integration of the corresponding expressions in \cite{Eliz11}, one gets (no
summation over $\mu $) $\left\langle T_{\mu }^{\mu }\right\rangle _{\mathrm{%
ren}}^{\mathrm{(M)}}=m\left\langle \bar{\psi}\psi \right\rangle _{\mathrm{ren%
}}^{\mathrm{(M)}}/D$, for $\mu =0,2,\ldots ,D$, and $\left\langle
T_{1}^{1}\right\rangle _{\mathrm{ren}}^{\mathrm{(M)}}=0$. These results
could be obtained from general arguments: from the covariant conservation
equation $\partial \left\langle T_{\mu }^{\alpha }\right\rangle _{\mathrm{ren%
}}^{\mathrm{(M)}}/\partial x^{\prime \alpha }=0$ and from the trace relation
$\left\langle T_{\mu }^{\mu }\right\rangle _{\mathrm{ren}}^{\mathrm{(M)}%
}=m\left\langle \bar{\psi}\psi \right\rangle _{\mathrm{ren}}^{\mathrm{(M)}}$%
. The effects of a brane with bag boundary condition and cosmic string on
the Dirac vacuum in background of (4+1)-dimensional anti-de Sitter spacetime
were studied in \cite{Bell22AdS}. It can be checked that the result (\ref%
{FCM}) for $D=4$ is obtained from the corresponding formula in Ref. \cite%
{Bell22AdS} in the flat spacetime limit and in the absence of a cosmic
string. For a massless field, from (\ref{FCM}) we get%
\begin{equation}
\left\langle \bar{\psi}\psi \right\rangle _{\mathrm{ren}}^{\mathrm{(M)}}=-%
\frac{N\Gamma \left( \frac{D+1}{2}\right) }{\left( 4\pi \right) ^{\frac{D+1}{%
2}}|x^{\prime 1}|^{D}},\;m=0,  \label{FCMm0}
\end{equation}%
and the VEV of the energy-momentum tensor vanishes $\left\langle T_{\mu
}^{\nu }\right\rangle _{\mathrm{ren}}^{\mathrm{(M)}}=0$. Comparing with the
results obtained above for the Fulling-Rindler vacuum, we see that the
situation here is significantly different from the case of a boundary
uniformly accelerating through the Fulling-Rindler vacuum. In the latter
case, the fermion condensate vanishes for a massless field in spatial
dimensions $D\geq 2$, while the VEV of the energy-momentum tensor is nonzero.

The above calculation of the VEVs does not use a specific representation of
the Dirac matrices $\gamma ^{(b)}$. In particular, this means that the
results obtained for the fermion condensate and the vacuum energy-momentum
tensor apply to both types of Dirac fields that realize two inequivalent
irreducible representations of the Clifford algebra in an odd number of
spacetime dimensions ($D$ being an even number). 2D Dirac materials are
among the interesting condensed matter realizations of the Dirac model in
spatial dimension $D=2$. The long-wavelength dynamics of the electronic
subsystem in such materials (the most famous example of which is graphene)
is described by the Dirac equation, in which the Fermi velocity of the
electrons is used instead of the speed of light (see, e.g., \cite%
{Gusy07,Cast09,Vafe14}). An effective Rindler metric for the corresponding
quasiparticles can be generated by engineering the strain profile in
graphene (see, for example, \cite{Iori12,Iori14,Bhar23} and references
therein). The setup we discussed can be used to model and investigate
edge-induced effects in these 2D structures within the framework of
effective Dirac theory.

\section{Fermionic Casimir densities in gravitational fields}

\label{sec:Grav}

In this section, we will apply the above results to the study of
boundary-imposed contributions to the local characteristics of the fermionic
vacuum in gravitational fields. The cases of weak gravitational fields and
background geometries that are conformally related to Rindler spacetime will
be discussed.

\subsection{Weak gravitational fields}

For a weakly gravitating system of a mass $M$, the line element in a $(D+1)$%
-dimensional spacetime is approximated by (see, for example, \cite{Empa08})
\begin{equation}
ds_{\mathrm{g}}^{2}\approx \left[ 1-w(r^{\prime })\right] dt^{\prime 2}-%
\left[ 1+\frac{w(r^{\prime })}{D-2}\right] \sum_{i=1}^{D}(dx^{\prime i})^{2},
\label{dsg}
\end{equation}%
in the coordinates $x^{\prime \mu }$. Here, $w(r^{\prime })=(r_{g}/r^{\prime
})^{D-2}\ll 1$, $r^{\prime 2}=\sum_{i=1}^{D}(x^{\prime i})^{2}$, and%
\begin{equation}
r_{g}^{D-2}=\frac{16\pi GM}{\left( D-1\right) \Omega _{D-1}},  \label{rg}
\end{equation}%
with $\Omega _{D-1}=2\pi ^{D/2}/\Gamma (D/2)$ being the area of a unit $%
(D-1) $-sphere. The acceleration at a given point is expressed as
\begin{equation}
g(r^{\prime })=\frac{D/2-1}{r^{\prime }}w(r^{\prime }).  \label{gr}
\end{equation}

Directing the local axis $x^{\prime 1}$ along the radial direction and
expanding the metric around $r^{\prime }=r_{0}^{\prime }$ one gets%
\begin{equation}
ds_{\mathrm{g}}^{2}\approx \left( 1-w_{0}+2g_{0}x^{\prime 1}\right)
dt^{\prime 2}-\left( 1+\frac{w_{0}-2g_{0}x^{\prime 1}}{D-2}\right)
\sum_{i=1}^{D}(dx^{\prime i})^{2},  \label{dsg2}
\end{equation}%
where $w_{0}=w(r_{0}^{\prime })$ and $g_{0}=g(r_{0}^{\prime })$. In the next
step, we make a transition to a new coordinate system $(t,z,\mathbf{x})$,
with $\mathbf{x}=\left( x^{2},\ldots ,x^{D}\right) $, defined by the
relations (for the transformation in the special case $D=3$ see also \cite%
{Sorg21})%
\begin{eqnarray}
t^{\prime } &=&\frac{t}{\sqrt{1-w_{0}}},\;x^{\prime 1}=\frac{z+\frac{%
g_{0}\left( z^{2}-\mathbf{x}^{2}\right) }{2\left( D-2\right) }}{\sqrt{1+%
\frac{w_{0}}{D-2}}},  \notag \\
x^{\prime i} &=&\frac{1+\frac{g_{0}z}{D-2}}{\sqrt{1+\frac{w_{0}}{D-2}}}%
x^{i},\;i\geq 2.  \label{xpi}
\end{eqnarray}%
This brings the line element to the form%
\begin{equation}
ds_{\mathrm{g}}^{2}\approx \left( 1+g_{0}z\right) ^{2}dt^{2}-dz^{2}-d\mathbf{%
x}^{2}.  \label{dsg3}
\end{equation}%
Introducing the coordinates $(\tau ,\rho )$ according to $\tau =g_{0}t$ and $%
\rho =z+1/g_{0}$, the line element is written in the Rindler form (\ref{ds2}%
).

Having the VEVs in the Rindler spacetime, the corresponding VEVs for a
boundary in the geometry (\ref{dsg}) are found by a coordinate
transformation $\left( \tau ,\rho ,\mathbf{x}\right) \rightarrow \left(
t^{\prime },x^{\prime 1},\mathbf{x}^{\prime }\right) $. We denote these VEVs
by $\left\langle \bar{\psi}\psi \right\rangle ^{\prime }$ and $\langle
T_{\nu }^{\mu }\rangle ^{\prime }$ for the fermion condensate and vacuum
energy-momentum tensor. The fermion condensate is a scalar and one gets $%
\left\langle \bar{\psi}\psi \right\rangle ^{\prime }=\left\langle \bar{\psi}%
\psi \right\rangle $. The VEV of the energy-momentum tensor is obtained by
using the standard law for the transformation of second rank tensors (no
summation over $\mu $):%
\begin{equation}
\langle T_{\mu }^{\mu }\rangle ^{\prime }=\langle T_{\mu }^{\mu }\rangle
,\;\langle T_{l}^{1}\rangle ^{\prime }=\frac{g_{0}x^{\prime l}}{D-2}\left(
\langle T_{1}^{1}\rangle -\langle T_{2}^{2}\rangle \right) ,\;l=2,\ldots ,D.
\label{Tmug}
\end{equation}%
In the absence of boundaries, the vacuum stresses are isotropic, $\langle
T_{1}^{1}\rangle _{0}^{\mathrm{ren}}=\langle T_{2}^{2}\rangle _{0}^{\mathrm{%
ren}}$, and the corresponding off-diagonal contribution in (\ref{Tmug})
vanishes. The appearance of the off-diagonal component is a purely
boundary-induced effect.

\subsection{Gravitational fields conformally related to the Rindler spacetime%
}

With the results presented above, we can find the boundary induced VEVs for
a massless Dirac field in the problems where the background geometry is
conformally related to the Rindler spacetime. Under the conformal
transformation $\tilde{g}_{\mu \nu }=\Omega ^{2}(x)g_{\mu \nu }$ of the
metric tensor the Dirac matrices and the field transform as $\tilde{\gamma}%
^{\mu }=\Omega ^{-1}\gamma ^{\mu }$ and $\tilde{\psi}=\Omega ^{-D/2}\psi $
and the Dirac equation for massless fields is conformally invariant. In the
presence of boundaries, by taking into account that the normal to the
boundary transforms as $\tilde{n}_{\mu }=\Omega n_{\mu }$, we conclude that
the bag boundary condition is also invariant under conformal
transformations. Consequently, by conformal transformation one can relate
the boundary-induced VEVs in conformally connected problems. For a massless
field, the fermion condensate becomes zero in spatial dimensions $D\geq 2$
and we will focus on the VEV of the energy-momentum tensor.

Let us denote by $\langle \tilde{T}_{\mu }^{\nu }\rangle $ the VEV of the
energy-momentum tensor for a fermionic field in a spacetime with the metric
tensor $\tilde{g}_{\mu \nu }$ conformally related to the metric tensor
defined by (\ref{ds2}). We assume the presence of a boundary which is a
conformal image of the boundary $\rho =\rho _{0}$ in the problem on the
Rindler bulk. The expectation value $\langle \tilde{T}_{\mu }^{\nu }\rangle $
is decomposed into two contributions. The first one, denoted here as $%
\langle \tilde{T}_{\mu }^{\nu }\rangle _{\mathrm{G}}$, is the geometrical
part and is uniquely determined by the geometrical characteristics of the
curved metric $\tilde{g}_{\mu \nu }$. The second contribution is conformally
related to the VEV $\langle T_{\mu }^{\nu }\rangle $ for the Fulling-Rindler
vacuum and the total VEV is expressed as (see, for example, \cite{Birr82})%
\begin{equation}
\langle \tilde{T}_{\mu }^{\nu }\rangle =\langle \tilde{T}_{\mu }^{\nu
}\rangle _{\mathrm{G}}+\frac{\langle T_{\mu }^{\nu }\rangle }{\Omega
^{D+1}(x)}.  \label{emtrel}
\end{equation}%
In this relation, it is assumed that the vacuum state in the problem with
the metric tensor $\tilde{g}_{\mu \nu }$ is the conformal counterpart of the
Fulling-Rindler vacuum. A similar relation holds for the boundary-free
parts, $\langle \tilde{T}_{\mu }^{\nu }\rangle _{0}=\langle \tilde{T}_{\mu
}^{\nu }\rangle _{\mathrm{G}}+\Omega ^{-D-1}\langle T_{\mu }^{\nu }\rangle
_{0}$. Combining this with (\ref{emtrel}), we get the relation%
\begin{equation}
\langle \tilde{T}_{\mu }^{\nu }\rangle _{\mathrm{b}}=\frac{\langle T_{\mu
}^{\nu }\rangle _{\mathrm{b}}}{\Omega ^{D+1}(x)}  \label{emtb}
\end{equation}%
for the boundary-induced VEVs.

As an example we consider a cosmological model with negative curvature
spatial foliation (open universe). In conformal time $\tau $ and spherical
spatial coordinates $(\chi ,\theta _{1},\theta _{2},\ldots ,\theta _{D-1})$,
with $0\leq \theta _{l}\leq \pi $, $l=1,\ldots ,D-2$, $0\leq \theta
_{D-1}\leq 2\pi $, the line element is given by
\begin{equation}
d\tilde{s}^{2}=a^{2}(\tau )\left( d\tau ^{2}-d\chi ^{2}-\sinh ^{2}\chi
d\Omega _{D-1}^{2}\right) ,  \label{dstilde}
\end{equation}%
where $a(\tau )$ is the scale factor. The synchronous time coordinate $t$ is
defined by $t=\int d\tau \,a(\tau )$. In order to bring the Rindler line
element (\ref{ds2}) into the form conformal to (\ref{dstilde}), we make
spatial coordinate transformation (see also \cite{Birr82} for the case $D=3$%
)
\begin{eqnarray}
\rho &=&\alpha _{0}\left( \cosh \chi -\sinh \chi \cos \theta _{1}\right)
^{-1},  \notag \\
x^{i} &=&\rho \sinh \chi \cos \theta _{i}\prod_{n=1}^{i-1}\sin \theta
_{n},\;x^{D}=\rho \sinh \chi \prod_{n=1}^{D-1}\sin \theta _{n},
\label{Trans1}
\end{eqnarray}%
with $i=2,\ldots ,D-1$ and $\alpha _{0}$ being a constant. With this
transformation, the line element becomes%
\begin{equation}
ds_{\mathrm{R}}^{2}=\rho ^{2}\left( d\tau ^{2}-d\chi ^{2}-\sinh ^{2}\chi
d\Omega _{D-1}^{2}\right) .  \label{ds2Conf}
\end{equation}%
From here, the conformal relation $d\tilde{s}^{2}=[a(\tau )/\rho ]^{2}ds_{%
\mathrm{R}}^{2}$ is seen with the cosmological metric defined by (\ref%
{dstilde}). The conformal factor in (\ref{emtb}) is given by $\Omega
(x)=a(\tau )/\rho $.

In the relation (\ref{emtrel}) the left- and right-hand sides are witten in
the same coordinate system. Considering the coordinates $(\tau ,\chi ,\theta
_{1},\theta _{2},\ldots ,\theta _{D-1})$, one needs to transform the
energy-momentum tensor $\langle T_{\mu }^{\nu }\rangle $ from Section \ref%
{sec:EMT} to this coordinate system. We will denote the transformed VEV by $%
\langle \tilde{T}_{\mu }^{\nu }\rangle _{\mathrm{R}}$. By using the standard
transformation law for the second rank tensors, the following expressions
are obtained for the diagonal components (no summation over $l$):%
\begin{eqnarray}
\langle \tilde{T}_{l}^{l}\rangle _{\mathrm{R}} &=&\langle T_{l}^{l}\rangle
,\;l=0,3,\ldots .D,  \notag \\
\langle \tilde{T}_{l}^{l}\rangle _{\mathrm{R}} &=&\langle T_{l}^{l}\rangle +%
\frac{(-1)^{l}\left( \langle T_{1}^{1}\rangle -\langle T_{2}^{2}\rangle
\right) \sin ^{2}\theta _{1}}{\left( \cosh \chi -\sinh \chi \cos \theta
_{1}\right) ^{2}},l=1,2.  \label{Tlltilde}
\end{eqnarray}%
In addition to the diagonal components, the vacuum energy-momentum tensor in
the new coordinate system also has a non-zero off-diagonal component%
\begin{equation}
\langle \tilde{T}_{2}^{1}\rangle _{\mathrm{R}}=\frac{1-\coth \chi \cos
\theta }{\left( \coth \chi -\cos \theta \right) ^{2}}\sin \theta \left(
\langle T_{1}^{1}\rangle -\langle T_{2}^{2}\rangle \right) .
\label{T12tilde}
\end{equation}%
As expected, the trace of the energy-momentum tensor is not changed under
the coordinate transformation.

In the problem without boundaries the vacuum stresses are isotropic, $%
\langle T_{1}^{1}\rangle _{0}=\langle T_{2}^{2}\rangle _{0}$, and we obtain $%
\langle \tilde{T}_{\mu }^{\nu }\rangle _{0\mathrm{R}}=\langle T_{\mu }^{\nu
}\rangle _{0}$. By using the result for $\langle T_{\mu }^{\nu }\rangle _{0}$
from \cite{Bell23} and the conformal relation (\ref{emtrel}), for the
boundary-free part one gets
\begin{equation}
\langle \tilde{T}_{\mu }^{\nu }\rangle _{0}=\langle \tilde{T}_{\mu }^{\nu
}\rangle _{\mathrm{G}}+\frac{N[a(\tau )]^{-D-1}}{2^{D}\pi ^{\frac{D}{2}%
}\Gamma \left( \frac{D}{2}\right) }\int_{0}^{\infty }d\omega \,\frac{\omega
^{D}B_{D}(\omega )}{e^{2\pi \omega }-\left( -1\right) ^{D}}\mathrm{diag}%
\left( -1,\frac{1}{D},\ldots ,\frac{1}{D}\right) ,  \label{emtt0}
\end{equation}%
where $B_{1}=B_{2}=1$ and
\begin{equation}
B_{D}(\omega )=\prod\limits_{l=1}^{[\frac{D-1}{2}]}\left[ 1+\left( \frac{l-\{%
\frac{D}{2}\}}{\omega }\right) ^{2}\right] ,  \label{BD}
\end{equation}%
for $D\geq 3$. In (\ref{BD}), $[\frac{D-1}{2}]$ and $\{\frac{D}{2}\}$
correspond to the integer and fractional parts of the enclosed numbers,
respectively. The spatial geometry under consideration is maximally
symmetric and as expected the VEV (\ref{emtt0}) is homogeneous. For the
boundary-induced part, from (\ref{emtb}) we have $\langle \tilde{T}_{\mu
}^{\nu }\rangle _{\mathrm{b}}=[\rho /a(\tau )]^{D+1}\langle \tilde{T}_{\mu
}^{\nu }\rangle _{\mathrm{bR}}$, where the boundary-induced contribution $%
\langle \tilde{T}_{\mu }^{\nu }\rangle _{\mathrm{bR}}$ is given by the
right-hand sides of (\ref{Tlltilde}) and (\ref{T12tilde}) with replacement $%
\langle T_{\mu }^{\nu }\rangle \rightarrow \langle T_{\mu }^{\nu }\rangle _{%
\mathrm{b}}$. The boundary-induced vacuum stresses are anisotropic, $\langle
T_{1}^{1}\rangle _{\mathrm{b}}\neq \langle T_{2}^{2}\rangle _{\mathrm{b}}$,
and the off-diagonal component is not zero. By taking into account that for
a massless field the combination $\rho ^{D+1}\langle T_{\mu }^{\nu }\rangle
_{\mathrm{b}}$ depends on $\rho $ and $\rho _{0}$ through the ratio $\rho
/\rho _{0}$ (see Eqs. (\ref{EMTRRm0}) and (\ref{EMTRLm0})), we see that the
same holds for the VEV $\langle \tilde{T}_{\mu }^{\nu }\rangle _{\mathrm{b}}$%
. Again, this is a consequence of the homogeneity of the background space.
The location of the boundary in the new coordinates is given by
\begin{equation}
\cosh \chi -\sinh \chi \cos \theta _{1}=\frac{\rho _{0}}{\alpha _{0}}.
\label{Bound}
\end{equation}%
In the theory of the Casimir effect, closed analytical results for vacuum
characteristics are usually obtained for boundaries that are coordinate
surfaces (e.g., flat, spherical, cylindrical boundaries). As seen from Eq. (%
\ref{Bound}), the geometry of the boundary in the example under
consideration is quite complicated in terms of the initial coordinates.

Let us consider some special cases of the scale factor in (\ref{dstilde}).
For the Milne universe, the scale factor in synchronous time is given by $%
a=t $. In terms of the conformal time $\tau $, it is expressed as $a(\tau
)=\alpha e^{\tau }$ with a constant $\alpha $. The Milne universe is flat
and $\langle \tilde{T}_{\mu }^{\nu }\rangle _{\mathrm{G}}=0$. Another
example corresponds to the de Sitter spacetime foliated by hyperbolic
coordinates. The corresponding line element is given by (\ref{dstilde}) with
$a^{2}(\tau )=\alpha ^{2}/\sinh ^{2}\tau $. In terms of the time $t$, the
scale factor becomes $a^{2}=\alpha ^{2}\sinh ^{2}(t/\alpha )$.

\section{Conclusion}

\label{sec:Conc}

We have studied the fermionic Casimir densities induced by a planar boundary
uniformly accelerated through the Fulling-Rindler vacuum state. The boundary
separates the right Rindler wedge into two regions: the region between the
Rindler horizon and boundary with $0<\rho <\rho _{0}$ (RL region) and the
region $\rho _{0}<\rho <\infty $ (RR region). On the boundary the field is
constrained by the bag boundary condition and we have found the complete set
of corresponding fermionic modes in both regions. The modes are specified by
the set of quantum numbers $(\omega ,\mathbf{k},\eta )$ where $\mathbf{k}$
is the momentum along the directions perpendicular to the direction of
acceleration and $\eta $ enumerates the spinorial degrees of freedom. The
eigenvalues of the energy $\omega $ are continuous in the RL region. In the
RR region, the energy spectrum is discrete and the corresponding eigenvalues
are roots of the equation (\ref{BCRR}). For the evaluation of the VEVs of
physical observables, bilinear in the field operator, the summation
technique over the complete set of modes is employed. The mode sums in the
RR region contain a summation over the eigenvalues $\omega _{n}=\omega
_{n}(\lambda )$ given implicitly. For the summation of the corresponding
series we use a variant of the generalized Abel-Plana formula. This allows
to separate the boundary-induced contributions in the VEVs. They are
presented in the form of integrals exponentially convergent for points away
from the boundary. Similar representations are obtained for the VEVs in the
RL region.

As important local characteristics of the fermionic vacuum, we have
considered the fermion condensate and the expectation value of the
energy-momentum tensor. They are splitted into boundary-free and
boundary-induced contributions. For points outside the boundary the
renormalization is required for the first parts only. The corresponding
renormalized VEVs for a massive field in general number of spatial
dimensions were investigated in \cite{Bell23}. The fermion condensate and
the vacuum energy-momentum tensor are expressed as (\ref{FC0r2}) and (\ref%
{Tmu0}). The vacuum stresses for the Fulling-Rindler vacuum in the
boundary-free problem are isotropic. That is not the case for the
boundary-induced contributions. For the fermion condensate, those
contributions are given by the expressions (\ref{FCRR3}) and (\ref{FCLR3})
in the RR and RL regions, respectively. The corresponding formulas for the
VEV of the energy-momentum tensor are given by (\ref{EMTRR3}) and (\ref%
{EMTRL4}) with the functions $F_{\mu }\left[ G_{\nu }\left( x\right) \right]
$ defined in (\ref{FRR1}). We have explicitly checked that the
boundary-induced parts in the VEVs obey the covariant conservation equation
and the trace relation (\ref{Trace}). The parallel stresses in the problem
on background of $(D+1)$-dimensional spacetime are connected with the
fermion condensate in $(D+3)$ dimensional spacetime by simple relations (\ref%
{relFC}).

For $D\geq 2$ the fermion condensate vanishes for a massless field. In the
case $D=1$, the condensate for a massless field in the RR region is given by
(\ref{FCD1m0}) and the corresponding expression for the RL region differs by
the sign. For a massive field, the boundary-free contribution to the fermion
condensate is negative. The boundary-induced part in the fermion condensate
is negative in the RR region and positive in the RL region. In the RR region
both the boundary-free and boundary-induced parts in the energy density and
in the effective pressures along the directions parallel to the mirror, $%
-\langle T_{l}^{l}\rangle $, $l=2,3,\ldots ,D$, are negative. In the RL
region these VEVs have opposite signs compared to the RR region. Near the
mirror, the total VEVs are dominated by the boundary-induced contributions,
whereas the boundary-free contributions are dominant near the Rindler
horizon. For a massive field, in the RR region and at large distances from
the mirror, $m\rho \gg 1$, the boundary-induced contribution dominates in
the range $m\rho _{0}>0.04$.

For a mirror at rest relative to an inertial observer in the Minkowski
vacuum the fermion condensate is given by (\ref{FCM}) and it is negative.
The normal stress vanishes and the parallel stresses are equal to the energy
density, determined from $\left\langle T_{0}^{0}\right\rangle _{\mathrm{ren}%
}^{\mathrm{(M)}}=m\left\langle \bar{\psi}\psi \right\rangle _{\mathrm{ren}}^{%
\mathrm{(M)}}/D$. For a massless field the VEV of the energy-momentum tensor
vanishes and for the fermion condensate one has the expression (\ref{FCMm0}%
). It would also be interesting to compare the above results with the
expectation values induced by a mirror moving with constant proper
acceleration in the Minkowski vacuum. However, no such investigation has
been conducted in the literature. When a mirror moves nonuniformly through
the Minkowski vacuum, real particles may be created. This phenomenon is
known as the dynamical Casimir effect (see, e.g., \cite{Bord09,Casi11,Dodo25}%
). The problem with general law of motion is exactly solvable for
conformally invariant fields in spatial dimension $D=1$. In higher
dimensions, different approximate methods have been used. In particular,
most investigations consider the nonrelativistic motion of mirrors,
including oscillatory motion.

In Section \ref{sec:Grav}, we use the results obtained for the Rindler
background spacetime to investigate the fermionic Casimir densities in
gravitational fields. Cases involving weak gravitational fields and
background geometries that are conformally related to Rindler spacetime are
discussed. One such application involves the consideration of the VEV of the
energy-momentum tensor in an open cosmological model described by the
Friedmann-Robertson-Walker metric. The corresponding conformal image of a
planar boundary in Rindler spacetime is rather complicated and the
boundary-induced energy-momentum tensor has also a nonzero off-diagonal
component.

\section*{Acknowledgments}

A.A.S. was supported by the grant No. 21AG-1C047 of the Higher Education and
Science Committee of the Ministry of Education, Science, Culture and Sport
RA. L.Sh.G. and V.Kh.K. were supported by the grant No. 21AG-1C069 of the
Higher Education and Science Committee of the Ministry of Education,
Science, Culture and Sport RA.

\appendix

\section{Normalization integral}

\label{sec:AppNI}

In this section we evaluate the integrals appearing in the normalization
conditions of the fermionic mode functions for the RR and RL regions. The
integrals have the form%
\begin{equation}
J_{\omega \omega ^{\prime }}=\chi _{\eta ^{\prime }}^{(j)\dagger }(\mathbf{k}%
)\int_{\rho _{1}}^{\rho _{2}}d\rho \,Z_{i\omega \prime +\frac{1}{2}\gamma
^{(0)}\gamma ^{(1)}}\left( \lambda \rho \right) Z_{i\omega -\frac{1}{2}%
\gamma ^{(0)}\gamma ^{(1)}}\left( \lambda \rho \right) \chi _{\eta }^{(+)}(%
\mathbf{k}),  \label{J1}
\end{equation}%
where the function $Z_{\nu }\left( x\right) $ is given by (\ref{ZIK}). By
using the relation (\ref{rel2}), we obtain the representation%
\begin{equation}
J_{\omega \omega ^{\prime }}=\frac{1}{2}\sum_{\varkappa =\pm 1}\chi _{\eta
^{\prime }}^{(j)\dagger }(\mathbf{k})\left( 1+\varkappa \gamma ^{(0)}\gamma
^{(1)}\right) \chi _{\eta }^{(+)}(\mathbf{k})\int_{\rho _{1}}^{\rho
_{2}}d\rho \,Z_{i\omega ^{\prime }+\frac{\varkappa }{2}}\left( \lambda \rho
\right) Z_{i\omega -\frac{\varkappa }{2}}\left( \lambda \rho \right) .
\label{J12}
\end{equation}%
The expression in the right-hand side is further simplified on the base of (%
\ref{RelComp}) and (\ref{rel22}):%
\begin{equation}
J_{\omega \omega ^{\prime }}=\frac{\delta _{\eta \eta ^{\prime }}}{2\lambda }%
\int_{\lambda \rho _{1}}^{\lambda \rho _{2}}dx\,\left[ Z_{i\omega ^{\prime }+%
\frac{1}{2}}\left( x\right) Z_{i\omega -\frac{1}{2}}\left( x\right)
+Z_{i\omega ^{\prime }-\frac{1}{2}}\left( x\right) Z_{i\omega +\frac{1}{2}%
}\left( x\right) \right] .  \label{J13}
\end{equation}

For the evaluation of the integral in (\ref{J13}) we note that the function $%
Z_{\nu }\left( x\right) $ obeys the equation for the modified Bessel
functions. From that eqaution it follows that (see also \cite{Prud2} for the
special case $Z_{\nu }\left( x\right) =K_{\nu }\left( x\right) $ and $%
x_{2}=\infty $)
\begin{equation}
\int_{x_{1}}^{x_{2}}\frac{dx}{x}\,Z_{\mu }(x)Z_{\nu }(x)=x\left. \frac{%
Z_{\mu }^{\prime }(x)Z_{\nu }(x)-Z_{\mu }(x)Z_{\nu }^{\prime }(x)}{\mu
^{2}-\nu ^{2}}\right\vert _{x=x_{1}}^{x=x_{2}}.  \label{Zint}
\end{equation}%
We will assume that the coefficients in (\ref{ZIK}) are taken in the way to
have the relation
\begin{equation}
Z_{\mu +1}(x)+Z_{\mu }^{\prime }(x)=\frac{\mu }{x}Z_{\mu }(x).  \label{Zrel}
\end{equation}%
In particular, this relation is obeyed for the functions $K_{\mu }(x)$ and $%
e^{-\mu \pi i}I_{\mu }(x)$. By using (\ref{Zrel}) in (\ref{Zint}) we obtain%
\begin{equation}
\int_{x_{1}}^{x_{2}}dx\,Z_{\nu }(x)Z_{\mu +1}(x)=\mu x\left. \frac{Z_{\mu
}^{\prime }(x)Z_{\nu }(x)-Z_{\mu }(x)Z_{\nu }^{\prime }(x)}{\mu ^{2}-\nu ^{2}%
}\right\vert _{x=x_{1}}^{x=x_{2}}-\int_{x_{1}}^{x_{2}}dx\,Z_{\nu }(x)Z_{\mu
}^{\prime }(x).  \label{Zint2}
\end{equation}%
From here it follows that%
\begin{equation}
\int_{x_{1}}^{x_{2}}dx\,\left[ Z_{\mu }(x)Z_{\nu +1}(x)+Z_{\nu }(x)Z_{\mu
+1}(x)\right] =x\left. \frac{Z_{\mu }(x)Z_{\nu +1}(x)-Z_{\mu +1}(x)Z_{\nu
}(x)}{\mu -\nu }\right\vert _{x=x_{1}}^{x=x_{2}}.  \label{Zint3}
\end{equation}

Now, in the general formula (\ref{Zint3}) we take $\mu =i\omega -1/2$ and $%
\nu =i\omega ^{\prime }-1/2$:%
\begin{equation}
\int_{x_{1}}^{x_{2}}dx\,\sum_{\varkappa =\pm 1}Z_{i\omega -\varkappa
/2}(x)Z_{i\omega ^{\prime }+\varkappa /2}(x)=\frac{x}{i}\sum_{\varkappa =\pm
1}\varkappa \left. \frac{Z_{i\omega -\varkappa /2}(x)Z_{i\omega ^{\prime
}+\varkappa /2}(x)}{\omega -\omega ^{\prime }}\right\vert
_{x=x_{1}}^{x=x_{2}}.  \label{Zint4}
\end{equation}%
With this result, for the integral appearing in the normalization conditions
one obtains
\begin{equation}
J_{\omega \omega ^{\prime }}=\delta _{\eta \eta ^{\prime }}\frac{\rho }{2i}%
\sum_{\varkappa =\pm 1}\varkappa \left. \frac{Z_{i\omega -\varkappa
/2}(\lambda \rho )Z_{i\omega ^{\prime }+\varkappa /2}(\lambda \rho )}{\omega
-\omega ^{\prime }}\right\vert _{\rho =\rho _{1}}^{\rho =\rho _{2}}.
\label{J14}
\end{equation}%
This result is used in the main text for the evaluation of the normalization
coefficients in the RR and RL regions.

\section{Summation formula over the eigenmodes in the RR region}

\label{sec:AppSum}

In this section we derive a summation formula over the eigenmodes $z=\omega
_{n}$ being the roots of the function $\bar{K}_{iz+1/2}(u)$. For that the
generalized Abel-Plana formula from \cite{SahaBook} will be used. In the
generalized Abel-Plana formula we take
\begin{align}
f(z)&=\frac{2}{\pi }\cosh \pi z\,F(z),  \notag \\
g(z)&=\frac{\bar{I}_{iz+1/2}(\eta )-\bar{I}_{-iz+1/2}(\eta )}{\bar{K}%
_{iz+1/2}(\eta )}F(z),  \label{fgAPF}
\end{align}%
where $F(z)$ is a function analytic in the right-half plane $\mathrm{Re}%
\,z>0 $ and the functions with bar are defined in (\ref{NotKbar}) and (\ref%
{NotIbar}). For the sum and difference of these functions one has%
\begin{equation}
g(z)\pm f(z)=\mp 2\frac{\bar{I}_{\mp iz+1/2}(\eta )}{\bar{K}_{iz+1/2}(\eta )}%
F(z).  \label{gpmf}
\end{equation}%
By taking into account that
\begin{equation}
\bar{I}_{-i\omega _{n}+1/2}(u)=-\bar{I}_{i\omega _{n}+1/2}(u),  \label{Iprop}
\end{equation}%
for the residues of the function $g(z)$ at the points $z=\omega _{n}$ we
obtain%
\begin{equation}
\mathrm{Res}_{z=z_{n}}g(z)=\frac{2\bar{I}_{iz+1/2}(u)}{\partial _{z}\bar{K}%
_{iz+1/2}(u)}F(z)|_{z=\omega _{n}}.  \label{Resg}
\end{equation}

Plugging (\ref{gpmf}) and (\ref{Resg}) in the generalized Abel-Plana formula
\cite{SahaBook}, the formula%
\begin{align}
& \lim_{h\rightarrow \infty }\left[ \sum_{n=1}^{n_{h}}\frac{i\bar{I}%
_{iz+1/2}(u)F(z)}{\partial _{z}\bar{K}_{iz+1/2}(u)}|_{z=\omega _{n}}-\frac{1%
}{\pi ^{2}}\int_{0}^{h}dx\,\cosh (\pi x)\,F(x)\right]   \notag \\
& \qquad =\frac{i}{2\pi }\int_{0}^{\infty }dx\,\frac{\bar{I}_{x+1/2}(u)}{%
\bar{K}_{x+1/2}(u)}\left[ F(xe^{\frac{\pi i}{2}})-F(xe^{-\frac{\pi i}{2}})%
\right] ,  \label{Sum}
\end{align}%
is obtained. Here, for a given $h$, the upper limit of the summation is
defined in accordance with $\omega _{n_{h}}<h<\omega _{n_{h}+1}$. From the
conditions on the functions $f(z)$ and $g(z)$ in the generalized Abel-Plana
formula we get the condition on the function $F(z)$ in (\ref{Sum}). For $%
z=x+iy$, with $x>0$, it takes the form%
\begin{equation}
|F(z)|<\varepsilon (|z|)e^{-\pi x}(|z|/u)^{2|y|},  \label{Fcond}
\end{equation}%
for large values of $|z|$, where $|z|\varepsilon (|z|)\rightarrow 0$ in the
limit $|z|\rightarrow 0$. From (\ref{Iprop}) it follows that the function $i%
\bar{I}_{i\omega _{n}+1/2}(\eta )$ is real. Note that we have the relation%
\begin{equation}
\bar{I}_{i\omega _{n}+1/2}\left( u\right) =-\frac{1}{uK_{i\omega
_{n}+1/2}\left( u\right) },  \label{IKbar}
\end{equation}%
and the function $iK_{i\omega _{n}+1/2}\left( u\right) $ is also real. Modes
of quantum fields expressed in terms of the modified Bessel function of the
second kind with an imaginary order appear also in other background
geometries. An example of a (2+1)-dimensional spacetime is the Beltrami
pseudosphere (see \cite{Saha24} for the corresponding vacuum currents). The
summation formula (\ref{Sum}) can be used to investigate edge-induced
effects in these geometries.

\section{Proof of the identities}

\label{sec:AppIdent}

In this section, the proof of the identity (\ref{Ident}) is presented. By
taking into account the relation (\ref{ZRL3}), we can write%
\begin{equation}
\frac{Z_{\nu _{\pm }}^{2}\left( u\right) }{\left\vert C_{\sigma }\right\vert
^{2}}=\frac{Z_{\nu _{\pm }}\left( u\right) }{C_{\sigma }}\left[ \frac{Z_{\nu
_{\mp }}\left( u\right) }{C_{\sigma }}\right] ^{\ast },  \label{Z2}
\end{equation}%
with $u=\lambda \rho $ and%
\begin{equation}
\nu _{\pm }=i\omega \pm \frac{1}{2}.  \label{nupm}
\end{equation}
Substituting the expression (\ref{ZRL2}) for the functions $Z_{i\omega \pm
\frac{1}{2}}\left( u\right) $ it can be seen that
\begin{align}
\frac{Z_{\nu _{\pm }}^{2}\left( u\right) }{\left\vert C_{\sigma }\right\vert
^{2}} &= K_{\nu _{\pm }}^{2}\left( u\right) -\bar{K}_{\nu
_{+}}(u_{0})\left\{ \frac{I_{\nu _{\pm }}\left( u\right) }{\bar{I}_{\nu
_{+}}(u_{0})}\frac{I_{-\nu _{\pm }}\left( u\right) }{\bar{I}_{-\nu
_{-}}(u_{0})}\bar{K}_{\nu _{+}}(u_{0})\right.  \notag \\
& \left. \pm \left[ \frac{I_{\nu _{\pm }}\left( u\right) }{\bar{I}_{\nu
_{+}}(u_{0})}-\frac{I_{-\nu _{\pm }}\left( u\right) }{\bar{I}_{-\nu
_{-}}(u_{0})}\right] K_{\nu _{+}}\left( u\right) \right\} ,  \label{Ident2}
\end{align}%
where $u_{0}=\lambda \rho _{0}$. Next, we use the relations%
\begin{align}
\bar{K}_{\nu _{+}}(z) &=-\pi \frac{\bar{I}_{-\nu _{-}}(z)+\bar{I}_{\nu
_{+}}(z)}{2\cosh \left( \omega \pi \right) },  \notag \\
K_{\nu _{\pm }}(u) &=\pm \pi \frac{I_{-\nu _{\pm }}(u)-I_{\nu _{\pm }}(u)}{%
2\cosh \left( \omega \pi \right) },  \label{RelK}
\end{align}%
to show that
\begin{equation}
\frac{Z_{\nu _{\pm }}^{2}\left( \lambda \rho \right) }{\left\vert C_{\sigma
}\right\vert ^{2}}-K_{\nu _{\pm }}^{2}\left( \lambda \rho \right) =\frac{\pi
\bar{K}_{\nu _{+}}(z)}{2\cosh \left( \omega \pi \right) }\left[ \frac{I_{\nu
_{\pm }}^{2}\left( \lambda \rho \right) }{\bar{I}_{\nu _{+}}(z)}+\frac{%
I_{-\nu _{\pm }}^{2}\left( \lambda \rho \right) }{\bar{I}_{-\nu _{-}}(z)}%
\right] .  \label{Ident0}
\end{equation}%
The identity (\ref{Ident}) is a direct consequence of this relation.

In order to evaluate the boundary-induced contribution in the VEV of the
energy density in the RL region, we need to have the analog of the identity (%
\ref{Ident0}) for the product $Z_{i\omega +1/2}\left( u\right) Z_{i\omega
-1/2}\left( u\right) $. In the way similar to that used for (\ref{Ident0}),
by using the relations (\ref{RelK}), it can be shown that%
\begin{equation}
\frac{Z_{\nu _{+}}\left( u\right) Z_{\nu _{-}}\left( u\right) }{\left\vert
C_{\sigma }\right\vert ^{2}}=K_{\nu _{+}}\left( u\right) K_{\nu _{-}}\left(
u\right) -\frac{\pi \bar{K}_{\nu _{+}}(u_{0})}{2\cosh \left( \omega \pi
\right) }\sum_{j=\pm 1}\frac{I_{ji\omega -\frac{1}{2}}(u)I_{ji\omega +\frac{1%
}{2}}\left( u\right) }{\bar{I}_{ji\omega +1/2}(u_{0})}.  \label{Ident3}
\end{equation}%
In order to obtain a similar identity for the function entering in the
expression of the stress $\langle T_{1}^{1}\rangle $, we use the relation%
\begin{equation}
Z_{\nu _{-}}\left( z\right) Z_{\nu _{+}}^{\prime }\left( z\right) -Z_{\nu
_{+}}\left( z\right) Z_{\nu _{-}}^{\prime }\left( z\right) =Z_{\nu
_{+}}^{2}(z)-Z_{\nu _{-}}^{2}(z)-2\frac{i\omega }{z}Z_{\nu _{+}}\left(
z\right) K_{\nu _{-}}(z),  \label{Ident4}
\end{equation}%
and the identities for the separate parts in the right-hand side, given
above.

\end{document}